\documentclass{aa}  
\usepackage{float}
\usepackage{enumitem}   
\usepackage{graphicx}
\usepackage{xcolor}
\usepackage{txfonts}
\usepackage{ulem}

\newcommand{\mgk}{\mbox{Mg\,{\sc ii}~k\,}}
\newcommand{\mghk}{\mbox{Mg\,{\sc ii}~h \&~k\,}}

\newcommand{\mg}{\mbox{Mg\,{\sc ii}~}}
\newcommand{\si}{\mbox{Si\,{\sc iv}~\,}}

\begin{document}

\title{Velocity field of an Active Region filament from GRIS IR He I and IRIS UV observations}

   \author{M. Murabito\inst{\ref{inst1},}\inst{\ref{inst2}}, V. Andretta\inst{\ref{inst3}}, S. Parenti\inst{\ref{inst4}}, C. Kuckein\inst{\ref{inst5},}\inst{\ref{inst6}}, S. J. González Manrique\inst{\ref{inst5},}\inst{\ref{inst6},}\inst{\ref{inst7}}, S. M. Lezzi\inst{\ref{inst3}}, S. L. Guglielmino\inst{\ref{inst8}}
   }

   \institute{INAF -- Osservatorio Astronomico di Roma, Via Frascati 33 I-00078 Monteporzio Catone, Rome Italy\\
    PDF \email{mariarita.murabitoi@inaf.it}\label{inst1}               \and
    Space Science Data Center (SSDC), Agenzia Spaziale Italiana, Via del Politecnico s.n.c., I-00133 Roma, Italy \label{inst2}    \and
    INAF -- Osservatorio Astronomico di Capodimonte, Salita Moiarello 16, I- 80131, Napoli, Italy \label{inst3} 
                   \and 
    Institut d’Astrophysique Spatiale, Bâtiment 121, Rue Jean Dominique Cassini, Université Paris Saclay, 91405 Orsay, France \label{inst4}
    \and
    Instituto de Astrof\'isica de Canarias (IAC), V\'ia L\'actea s/n, E-38205 La Laguna, Tenerife, Spain   \label{inst5}
    \and
    Departamento de Astrof\'\i{}sica, Universidad de La Laguna, E-38206 La Laguna, Tenerife, Spain \label{inst6}
        \and
    Institut für Sonnenphysik (KIS), Georges-Köhler-Allee 401a, 79110 Freiburg, Germany \label{inst7}
        \and
    INAF -- Osservatorio Astrofisico di Catania, Via Santa Sofia 78, I-95123 Catania, Italy \label{inst8}
    }

   \date{}

 
  \abstract
   {{Plasma flow measurements in solar active region filaments are rare, particularly in the infrared and ultraviolet ranges that probe the chromosphere and transition region. In addition, previous studies generally focused on prominences and filaments near the solar limb.} 
    } 
   {{This study presents a multi-wavelength, multi-instrument analysis of an active region filament observed on the solar disk on November 9 and 10, 2020. Our goal is to characterize the plasma flows in the filament using spectroscopic measurements in both the infrared and ultraviolet spectral ranges. This is important for understanding the mechanisms for filament support, mass loading, and energy balance. Furthermore, this also offers observational benchmarks for filament modeling and simulations.
   }}
   {Spectra from the IRIS satellite, including the \mgk 2796 \AA, \ion{C}{ii} 1335 \AA\ and \si 1393 \AA\ lines were analyzed alongside ground-based observations from the GREGOR Infrared Spectrograph and High-resolution Fast Imager instruments whose observed spectral ranges include the chromospheric \ion{He}{i} 10830 \AA\ and H$\alpha$ {6563} \AA\ lines. }
   {{Persistent blueshifts were measured within the filament structure in both spectral ranges. These can be interpreted as upflow velocities ranging from 0.5 to 15 km s$^{-1}$, with the \si 1393 \AA\ showing the highest values. Red shifted emission in the \ion{He}{i} and \mgk$_3$ at the footpoints of a newly formed dark bundle suggest chromospheric downflows, likely due to spatial overlap between an arch filament system close to the filament footpoints. The weak redshifted signal in the \si emission may suggest confinement to lower atmospheric layers.}
    }
   {{The observed velocity patterns provide, for the first time, a comprehensive and coherent view of the plasma dynamics from the chromosphere to the transition region, illustrating that the filament emission is consistently blueshifted in all the spectral windows, and thus in different temperature regimes.}
   }

   \keywords{Sun: Transition region - chromosphere - filaments - Sun: UV radiation -
             }
               
   \titlerunning{Multi-instrument study of plasma flow in an active region filament}
   \authorrunning{M. Murabito et al.}
   
   \maketitle
%

\section{Introduction}
\label{sec1}

Filaments and prominences are cool, dense structures of chromospheric plasma suspended in the corona above polarity inversion lines (PILs) in the photospheric magnetic field \citep{Parenti2014}. These structures are referred to as filaments when observed on the solar disk, and as prominences when seen off-limb. 
They are observed both in active regions and in the quiet Sun, where they are referred to as active region (AR) and quiescent filaments, respectively. AR filaments are typically more dynamic, smaller in size, and short-lived compared to their quiescent counterparts. Both quiescent and AR filaments form within so-called filament channels, which appears as a long, narrow structure characterized by a pattern of fibrils along the PILs. During the formation process, these fibrils gradually reorient from a perpendicular to a parallel alignment with respect to the PIL, over a timescale ranging from three hours to one or two days \citep{Gaizauskas1997,Wang_muglach2007}. 

Filament (and prominence) plasmas have temperature below 10$^{6}$ K and they are embedded in hotter coronal plasma. The interface between the prominence and the corona is known as the Prominence-Corona-Transition Region (PCTR). This is defined as the region where the plasma temperature rises from about 7000 K to 10$^{6}$ K in analogy to the transition region between the chromosphere and the corona. Different diagnostic tools are employed to study the two components of a filament: the cooler core and the PCTR. The former is investigated through hydrogen and helium spectral and continuum emissions. The latter, instead, is analyzed by observing transition-region spectral lines. Like in the filament core, the PCTR is also characterized of a variety of flows which are best measured in the ultraviolet (UV) and extreme UV spectral ranges.

Filamentary-like structures are also observed in emerging flux regions, which often display so-called arch filament systems \citep[AFS; e.g.,][]{Bruzek1967,Bruzek1969}. These structures consist of magnetic loops filled with cool plasma connecting opposite emerging polarities \citep[][]{Solanki2003}. They are typically observed in chromospheric lines such as H$\alpha$, Ca II infrared, Ca II H \& K, and \ion{He}{i} 10830 \AA\ \citep[e.g.,][]{Spadaro2004,Zuccarello2005,Murabito17,Sergio18,Sergio20,Reetika2024}, and exhibit upflows at loop tops and downflows at the footpoints of several tens of km s$^{-1}$ \citep[][]{Balthasar16,Zhong2019}.
quiescent and AR filaments, instead, are characterized by different kinds of flows. This has implications in our understanding the  mechanism of formation and their stability \citep{Parenti2014}. However, measurements of plasma flows in filaments are limited, and most of the available observations focus on quiescent filaments.

A few studies investigated AR filaments. For instance, using infrared (IR) photospheric \ion{Si}{i} and chromospheric \ion{He}{i} spectral lines, \citet{Kuckein2012b} reported small photospheric upflows of about $-0.15$ km s$^{-1}$ under an AR filament. The chromospheric portion of the same filament showed an average upward motion of $-0.24$ km s$^{-1}$. {H$\alpha$} observations reported by \citet{Ioshpa1999} show that AR filaments are located above regions characterized by chromospheric upflow motions, surrounded by areas of downflow motion. This property is shared with quiescent filaments.

Until the launch of the Interface Region Imaging Spectrograph \citep[IRIS,][]{depontieu2014}, most observations of filaments in the UV range had been obtained with the Ultraviolet Spectrograph and Polarimeter (UVSP) instrument on the Solar Maximum Mission \citep[SMM,][]{SMM1980} first, and later with the Solar Ultraviolet Measurements of Emitted Radiation \citep[SUMER,][]{SUMER1995} instrument aboard SOHO. These earlier studies outlined the following key findings: 
\begin{enumerate}[label=(\roman*)]
    \item persistent upflows 
    of 5.6 km s$^{-1}$ were observed at the filament locations using the SMM \ion{C}{iv} 1548 \AA\ line, formed in the PCTR at $10^{5}$ K \citep{Schmieder1984};
    \item a transition between redshifts and blueshifts was detected on either side of the filament using the SMM \ion{C}{iv} {1548} \AA\ and \si 1393 \AA\ spectral lines \citep{Engvold1985}. These results are consistent with the findings of \citet{Ioshpa1999}. \citet{Engvold1985} also reported velocity amplitude of $\pm$15 km s$^{-1}$ for both quiescent and AR filaments;
    \item redshifted brighter areas, interpreted as being near the filament footpoints, were reported using \si SUMER data, although no clear velocity signature was associated with the filament \citep{Kucera1999}.
\end{enumerate}

Doppler line shifts, and the corresponding line-of-sight (LoS) velocities, are affected by various sources of uncertainties. The inferred velocity values are position-dependent on the solar disk, and most of the reported measurements refer to filaments approaching the limb. This makes it difficult a direct comparison, as photospheric lines typically exhibit much smaller velocity shifts compared to those in the chromosphere or PCTR. A reliable zero-reference for the rest wavelength is ideally obtained using photospheric lines. However, such lines are rarely detectable in the UV spectral range. Therefore, setting an accurate zero-reference wavelength is necessary. 

Several studies have reported difficulties in achieving absolute wavelength calibration \citealp[e.g.,][]{Engvold1985,Kucera1999}. In the literature, velocity estimates are often derived under certain assumptions, such as averaging the wavelength over a full raster scan \citep{Brooks2009} or using quiet Sun (QS) regions far from the target as a reference \citep{DelZanna2009}. 
These methods generally result in uncertainties greater than 5 km s$^{-1}$ \citep{Winebarger2013}. A more accurate approach, though only feasible in the presence of two or more telluric lines has been used by \citet{Martinez1997} and \citet{Kuckein2012b}. As discussed in Section \ref{sec_gris}, this method achieves an uncertainty of less than 0.06 km s$^{-1}$. 
 
Given the observational issues described above, simultaneous high spectral and spatial resolution measurements of plasma flows in filaments, at various wavelengths, sampling different atmospheric layers, remain relatively rare. In this study, we present a multi-wavelength analysis of the velocity field in an AR filament, tracing plasma motions from the chromosphere up to the transition region. We use the most accurate velocity calibration to date for the employed spectral ranges.

\begin{figure*}
\includegraphics[scale=0.71,clip,trim=0 0 0 
0]{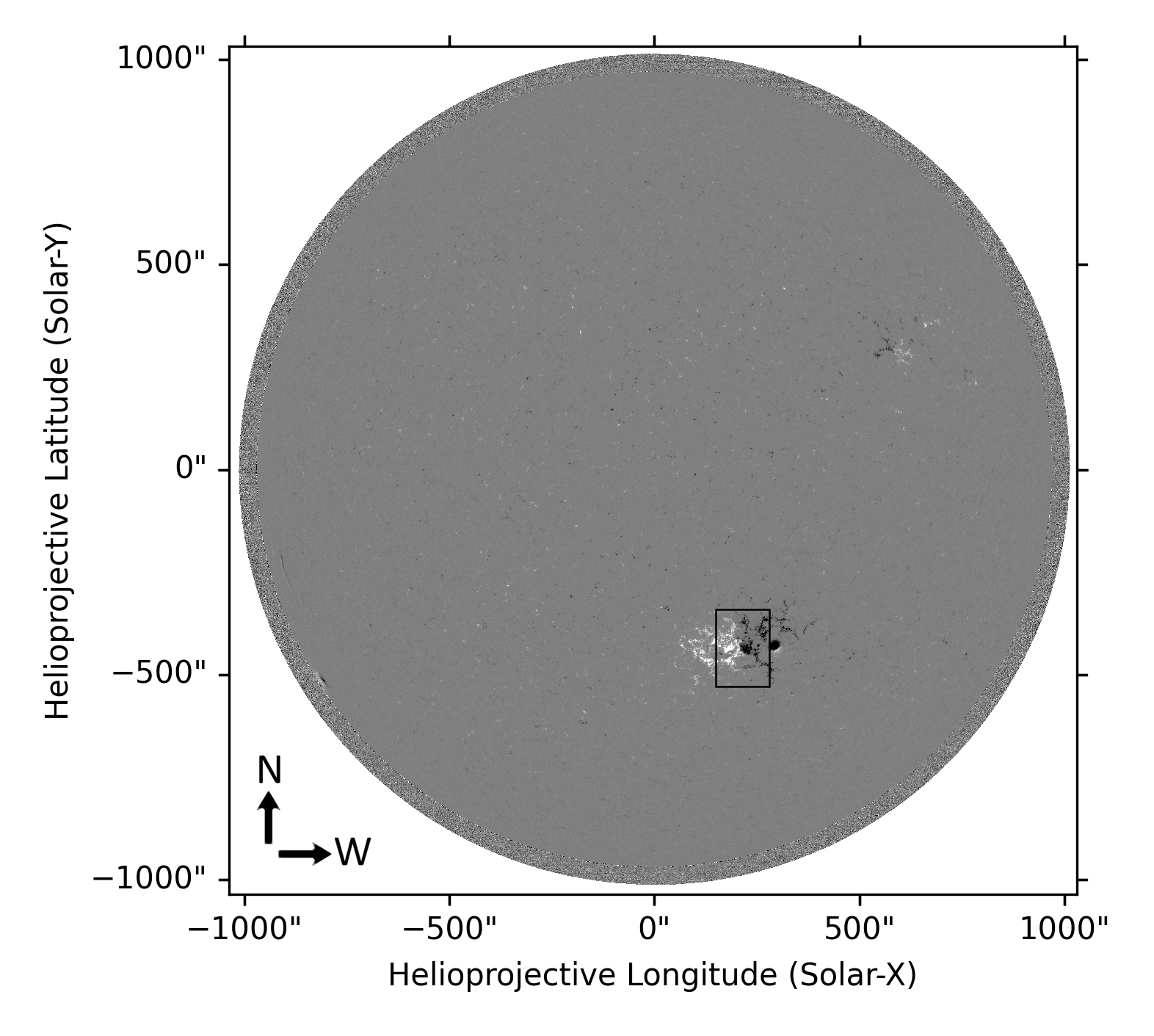}
\includegraphics[scale=0.42,trim=0 110 170 180]{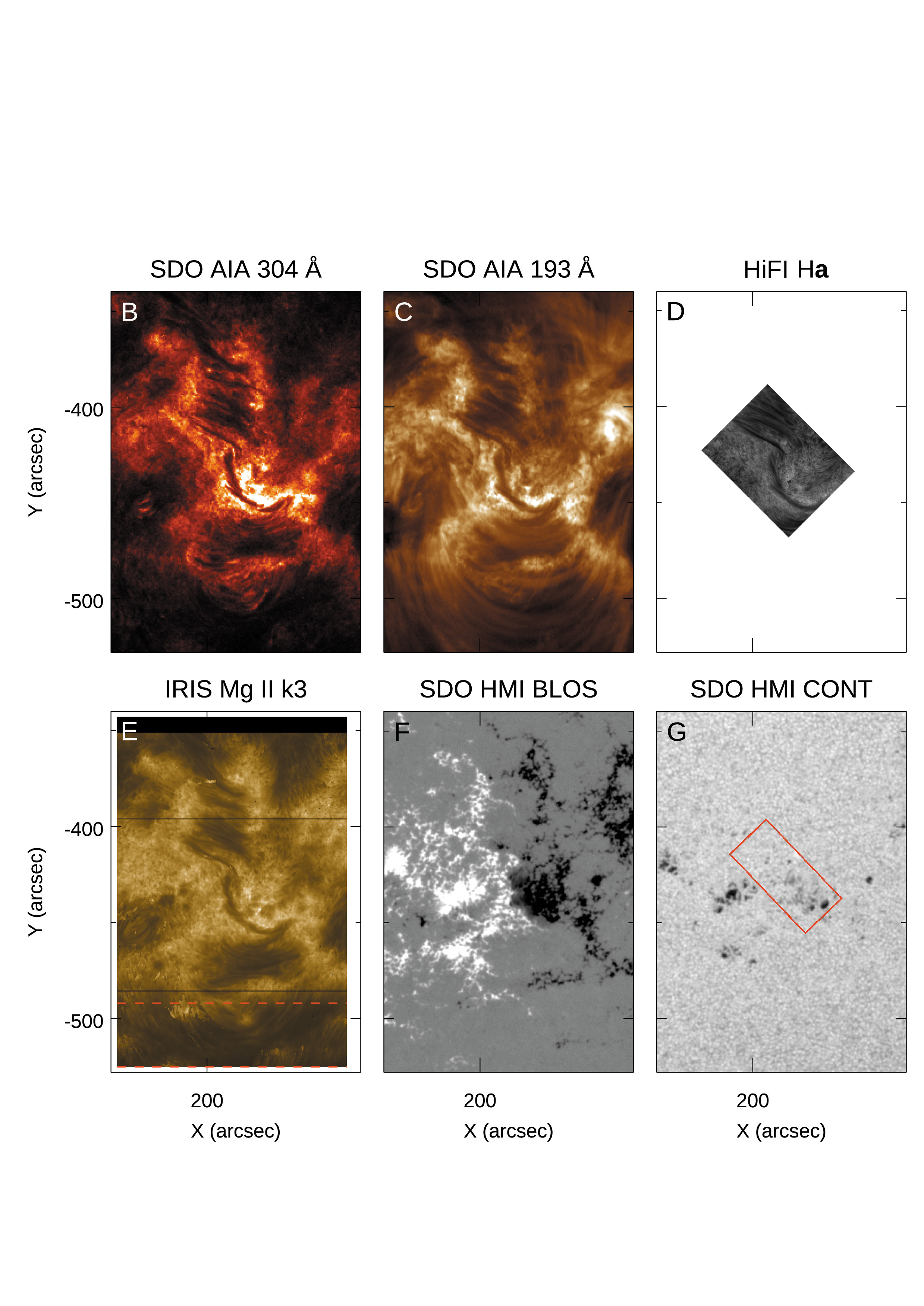}
\caption{Overall view of the AR12871. Panel A: Full-disk magnetogram clipped between $\pm$500 G from HMI taken at 10:55 UT on November 10. The black box indicates the FoV shown in the following AIA 304 and 193 \AA\ bands (panels B and C). Panel D: High-spatial resolution HiFI H$\alpha$ image taken at 10:48 UT on November 10. Panel E: IRIS slit-reconstructed image in the center of the \ion{Mg}{ii} k$_{3}$ line acquired between 11:03 UT and 11:52 UT on November 10. Panels F-G: HMI magnetogram clipped between $\pm$500 G and continuum taken at 10:55 UT. The red box in the continuum image indicates the GRIS FoV.}
\label{Fig1}
\end{figure*}

\section{Observations}

The AR filament analyzed in this study was observed on 2020 November 9 and 10 with the 1.5-meter GREGOR solar telescope \citep{Schmidt2012,Kleint2020} located on Tenerife island, Spain, during an observing campaign supported by the SOLARNET Transnational Access and Service Programme. The target (shown in the full-disk image of Fig. \ref{Fig1}A) was AR NOAA 12781, which exhibited a $\beta \gamma$ magnetic configuration consisting of a large sunspot in the leading negative polarity, an orphan penumbra, and several small pores in the following positive polarity (see Fig. \ref{Fig1}G). On November 9 and 10 the AR was located at solar heliocentric coordinates [X,Y]=[140\arcsec, -431\arcsec] and [250\arcsec, 427\arcsec], respectively. Figure \ref{Fig1} (panels B-G) displays the filament and its surrounding environment in AIA/SDO, HiFI/GREGOR and IRIS data. 
The filament is easily distinguishable in panels B, C, D and E as a compact and dark structure located between two AFSs, connecting the two opposite polarities seen in the magnetograms to the north and to the south of the filament itself. Surrounding the filament, particularly on its western side, intense activity is visible in terms of brightening events at chromospheric and coronal heights (Fig. \ref{Fig1} B and C). In contrast, the UV \mg k$_{3}$ intensity (Fig. \ref{Fig1} panel E) does not show any significant brightness enhancement in the same area.

The GREGOR observations consisted of full-Stokes spectropolarimetric data obtained with the GREGOR Infrared Spectrograph \citep[GRIS,][]{Collados2012} and of fast-imaging data obtained with the High-resolution Fast Imager \citep[HiFI,][]{Kuckein2017,Denker2023}. The adaptive optics system \citep{Berkefeld2012} was used to minimize the image degradation due to seeing effects. Several spectropolarimetric maps were taken on November 9 and 10. Here, we only use the scans acquired on November 10 between 09:17:32 and 09:37:05 UT, when the slit was aligned with the filament and scanned a large area of 56.3\arcsec$\times$10.4\arcsec. The spectral region observed with GRIS comprises the photospheric \ion{Si}{i} 10827 \AA\ and \ion{Ca}{i} 10839 \AA\ lines, as well as the chromospheric \ion{He}{i} 10830 \AA\ triplet, among other weaker lines. The step size of the slit was 0\farcs13 and each step consisted of 10 accumulations with an exposure time of 100 ms each. The spatial sampling along the slit was 0\farcs13. The field of view (FoV) is shown in Fig. \ref{Fig1} (panel G). A spatial scan consists of 200 steps in total scanned in about 20 minutes. In addition, the HiFI instrument recorded images of the target using two synchronized cameras, one observing the H$\alpha$ line core (Fig. \ref{Fig1}D) and the other the continuum close to H$\alpha$. These data were recorded in bursts of 500 images with a frame rate of 99\,Hz, from 08:44 UT to 11:02 UT on November 9 and from 08:44 UT to 10:49 UT on November 10 under good seeing conditions. 

The \ion{He}{i} 10830 \AA\ line is a widely used diagnostic of the chromosphere and transition region, sensitive to plasma dynamics across a range of temperatures, from the cool core of prominences and filaments to their PCTR. Its formation involves a complex interplay between photoionization-recombination processes, collisional excitation, and scattering of photospheric radiation by helium atoms in the metastable triplet state ($1s2s\ ^3S$). The line consists of three components, with rest wavelengths at 10830.34 \AA, 10830.25 \AA, and 10829.09 \AA. Unlike UV lines, these spectra can be reliably wavelength-calibrated, making \ion{He}{i} 10830 \AA\ a valuable reference for absolute velocity measurements.

We complement the study using the Atmospheric Imaging Assembly \citep[AIA,][]{lemen2012} and the Helioseismic and Magnetic Imager \citep[HMI,][]{scherrer2012} continuum filtergrams and magnetograms, 
aboard the Solar Dynamics Observatory \citep[SDO,][]{pesnell2012}. 
The HMI continuum filtergrams and magnetograms have been used to align all available data and locate the position of the filament with respect to the underlying photospheric magnetic field, while the AIA at 304 and 193 \AA\ images to study the upper atmospheric response.
The region of interest (RoI) was also observed by IRIS with many different observing modes and at many times during its passage across the disk (as for example Fig. \ref{Fig1}E). The different observing modes mainly concern the size of the area scanned from the slit. We analyzed in detail the spectra taken during three time intervals with the same observing mode, a dense 320-step raster that lasted approximately 50-60 minutes between November 9 and November 10, as reported in Table \ref{table1}.
All data analyzed in this work are listed in Table \ref{table1}.

\section{Data processing}

\subsection{Fast imaging data}\label{FastI}
The standard image-reduction steps were performed using the pipeline \texttt{sTools} \citep{Kuckein2017}, which includes dark-current and flat-field corrections, as well as image selection and alignment between the two cameras. The H$\alpha$ and H$\alpha$-continuum pairs of images were used for image restoration using the Multi-Object Multi-Frame Blind Deconvolution technique \citep[MOMFBD,][]{vannoort2005}. The best 50 frames of each burst were sufficient to obtain one restored image, 
covering a region of about $64\arcsec \times 40\arcsec$, or $1336 \times 1016$ pixels, with an image scale of 0\farcs048/pixel. 

\subsection{He 10830 \AA\ and Si 10827 \AA\ }\label{sec_gris}

Dark and flat-field corrections, as well as polarimetric calibration, were applied to the GRIS data using the standard data-reduction pipeline onsite \citep{Collados1999,Collados2003}. The GREGOR polarimetric unit was used for the polarimetric calibration \citep{Hofmann2012}. The Stokes profiles were normalized to the mean continuum, calculated from various QS areas within the FoV, excluding regions displaying notable polarization signal. An absolute wavelength calibration was possible using the two telluric H$_{2}$O lines within the observed spectral range. The obtained wavelength vector was corrected for Earth's orbital motions, solar rotation, and gravitational redshift and spanned between 10823.3 and 10841.4\,\AA\ \citep[for more details, see][]{Kuckein2012b}. 
The He I line-core slit-reconstructed image was obtained by calculating the minimum of the intensity within the He I spectral window.

The HAnle and ZEeman Light v2.0 code \citep[HAZEL2,][]{AsensioRamos2008}\footnote{The HAZEL2 code can be accessed at \url{https://github.com/aasensio/hazel2}} was utilized to perform inversions on the \ion{He}{i} triplet, as well as on \ion{Si}{i} and telluric lines. By simultaneously inverting these nearby lines, we can account for their wing overlaps caused by the close proximity of the \ion{Si}{i} line to the \ion{He}{i} triplet. Although the results for the \ion{Si}{i} line in the photosphere are not shown here, this process ensures a comprehensive approach. HAZEL2 is designed to incorporate various effects on the \ion{He}{i} triplet, including atomic level polarization, and Paschen-Back, Hanle, and Zeeman effects. In parallel, the \ion{Si}{i} line inversion is executed using the Stokes
Inversion based on Response functions (SIR) code \citep{ RuizCobo1992}, while the telluric line is fitted separately with a Voigt profile. The data derived from HAZEL2 ultimately provide insights into both the photosphere, derived from the \ion{Si}{i} 10827 \AA{} line, and the chromosphere, via the \ion{He}{i} 10830 \AA{} line. \\

For the \ion{He}{i} triplet inversion, HAZEL2 assumes a cloud model configuration, where the atmospheric parameters are kept constant within a slab located above the solar surface. The Stokes profiles for each pixel were fitted using a model with two components: one reflects the derived atmosphere, while the other represents stray light. A consistent stray-light profile was used across all spatial locations and times in the temporal series obtained as an average over a QS area near the observed pores. The Doppler velocity from the \ion{He}{i} triplet presented in this study was computed by inverting the full Stokes spectra with HAZEL2.

\subsection{UV IRIS data}
\label{UVproc}
The IRIS data were reduced and calibrated using the Solar Software (SSW) packages. These account for geometrical, flat-field, and dark current effects and for the orbital variation of the wavelength array as described in \citet{Wulser2018}. Furthermore, we use all the spectral lines in the NUV spectra to check the absolute wavelength calibration. 

We focus on the analysis of observations in the \mg (log$_{10}$T=4.15 K), \ion{C}{ii}~1335 \AA\ (log$_{10}$T=4.4 K) and \si doublet \AA\ (log$_{10}$T=4.9 K) spectral lines. These cover from the upper photosphere to the transition region, respectively \citep{Rathore2015a,Dufresne2021}.

We derived the most used spectral characteristics from the \mg lines, namely the k$_{1}$ (h$_{1}$), k$_{2}$ (h$_{2}$) and k$_{3}$ (h$_{3}$) features \citep{leenaarts2013}. 
These were extracted from the spectra using the automated procedure iris\_get\_mg\_features\_lev2 available in the SSW IRIS package, described in \citet{Pereira2013}. 
The error estimated for the k$_{3}$/h$_{3}$ Doppler velocity is the spectral sampling itself (0.05 \AA). Here, we report the intensity and Doppler velocity of the line core for the \mgk spectral line only, since the \mg h line behaves similarly.
Then, we used these quantities to estimate the k$_{2}$ peak difference (diff), k$_{2}$ peak separation (sep), sensitive to the upper chromospheric velocity gradient \citep{leenaarts2013}, and the k$_{2}$ peak asymmetry (asym), that gives the sign of the velocity above the $\tau =1$ level. The method to derive the diff, sep, and asym quantities is explained in the section \ref{mgd} of the Appendix.

We perform an absolute wavelength calibration for the \si and \ion{C}{ii} spectral lines. From the calibrated spectra, we derived the peak intensity, Doppler velocity and FWHM for the \si and \ion{C}{ii} 1335 \AA\ lines. For both lines, we perform a single Gaussian fit using the curvefit.pro routine from IDL. We discarded from the calculation pixels whose peak intensities were lower than 30 counts. In addition, we also restricted the computation to those profiles that have spectral widths larger than 70 m\AA. In the case of \si, we only show these calculated quantities for the \si 1393 \AA\ component. Regarding the \si velocity, the associated uncertainty is 1.1 km s$^{-1}$, as detailed in the section~\ref{si} of the Appendix. For QS regions, we find an average velocity value of 5.4 $\pm$ 0.6 km s$^{-1}$, slightly lower than that values reported in earlier studies, such as \citet{Chae1998}, \citet{Teriaca1999} and \citet{Peter1999}. All the details can be found in the section \ref{si} of the Appendix.

For \ion{C}{ii} 1335~\AA, although line profiles may exhibit single or double peaks due to temperature gradients between the chromosphere and the transition region \citep{Rathore2015a}, we adopted a single Gaussian fit for consistency, following similar approaches in previous studies \citep[e.g.,][]{Upendran2021}. This choice is based on multiple tests, which showed that in cases where a double-Gaussian fit is feasible, the resulting difference in velocity does not exceed 30\%. However, due to the generally noisy profiles and the estimated uncertainty of 3.3~km~s$^{-1}$ (see Appendix \ref{si}), we consider the \ion{C}{ii} Doppler velocities as qualitative indicators only.

\subsection{Co-alignment}

The alignment of all the available observations required several steps. In particular, HMI continuum images taken at the same time of the ground-based observations were used as a reference to scrutinize the magnetic polarities, with an accuracy of 0.5\arcsec. These were also used to align the H$\alpha$ broad- and narrow-band images. A finer alignment was necessary for all ground-based H$\alpha$ images, which were interpolated to have a common FoV and a common spatial scale. Then, the high-resolution H$\alpha$ image at 10:48 UT on November 10 was used as a reference to co-align the UV IRIS lines observations. This choice reflects the almost co-temporal time acquisition of the ground-based measurements and IRIS second scan. Finally, using the broad-band images (closest in time) we aligned the GRIS data.

\begin{table*}[]
\caption{Set of observations from different telescopes used in this study.}
\center
\label{table1}
\begin{tabular}{l l l l l l}
\hline 
\hline
                     &                       &                           &                         &                        &             \\

{Telescope}   & {Instrument}   & {Spectral coverage}& {Date}  & {Time coverage} &{FoV}     \\ 
                     &                       &                           &       [Day]             &    [Hour]              &  [arcsec]          \\
                     &                       &                           &                         &                        &                 \\                        
\hline
                     &                      &                            &                         &                        &                 \\
IRIS                 &                      &   \mg, \ion{C}{ii} and \si &       2020-11-09        &  20:26 - 21:15 UT      & 320 $\times$ 175  \\
                     &                      &   \mg, \ion{C}{ii} and \si &  2020-11-10             &   11:03 - 11:52 UT     & 320 $\times$ 175  \\
                     &                      &   \mg, \ion{C}{ii} and \si &  2020-11-10             &   20:48 - 21:37 UT     & 320 $\times$ 175  \\
                     &                      &                            &                         &                        &                      \\
SDO                  &  HMI                 & Fe I 6173 \AA\             & 2020-11-09              & 08:42 UT               & 70 $\times$ 60  \\
                     &                      &                            & 2020-11-10              & 08:42 UT               & 70 $\times$ 60             \\
                     &                      &                            &                         &                        &                           \\
GREGOR               & HiFI                 & H$\alpha$                  &   2020-11-09            & 08:44 UT - 11:02 UT    & 64 $\times$ 49    \\
                     &                      &                            &   2020-11-10            &  08:44 - 10:49 UT      & 64 $\times$ 49      \\
                     &                      &                            &                         &                        &                                \\
GREGOR               & GRIS                 & He I 10830 \AA\            &   2020-11-10            &  09:17 - 09:37 UT      & 56.3 $\times$ 10   \\
                     &                      &                            &                         &                        &                               \\
\hline
\end{tabular}
\end{table*}

\section{Results}

\begin{figure*}
\centering
\includegraphics[scale=0.85,clip,trim=0 410 0 280]{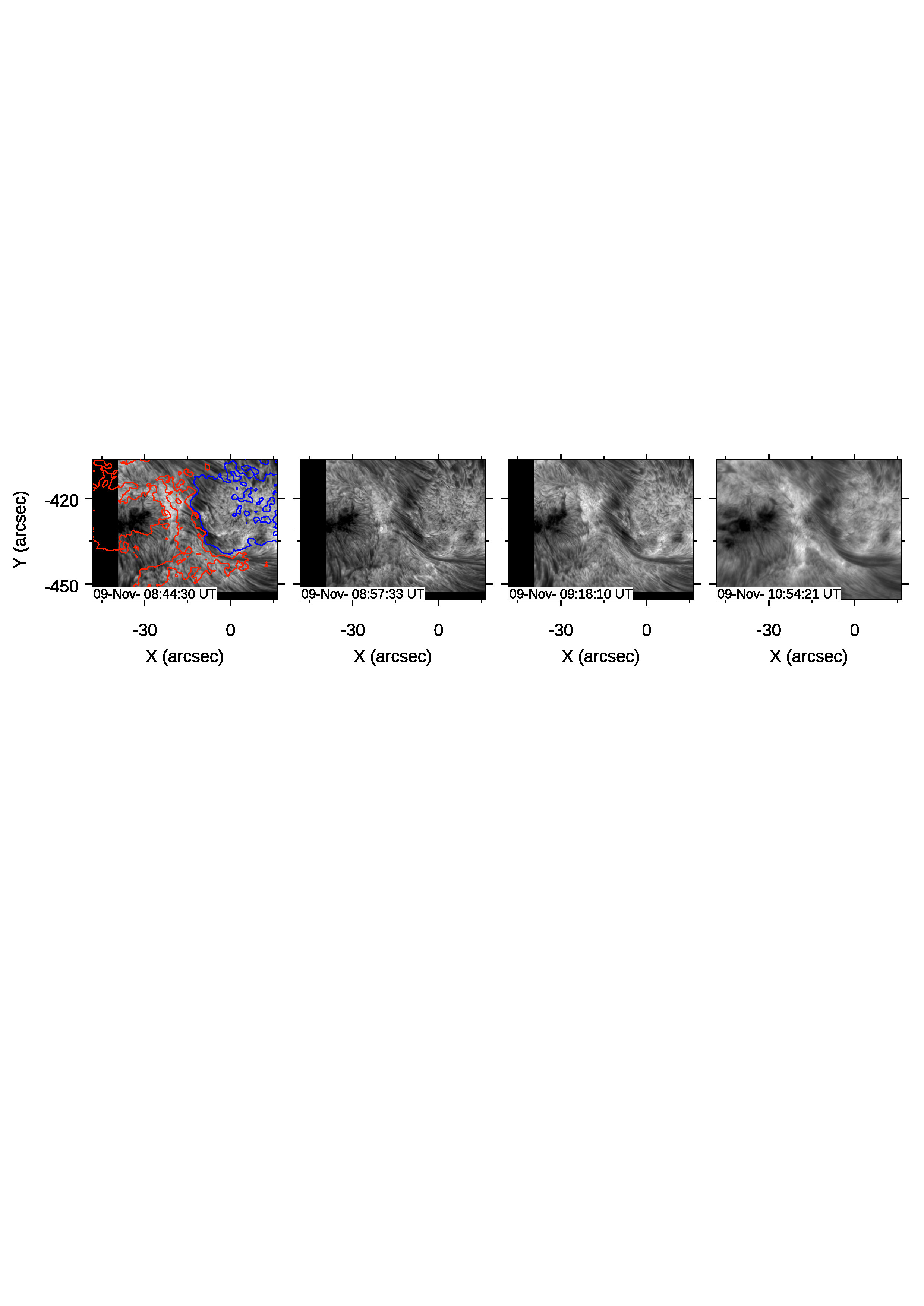}\\
\includegraphics[scale=0.85,clip,trim=0 400 0 270]{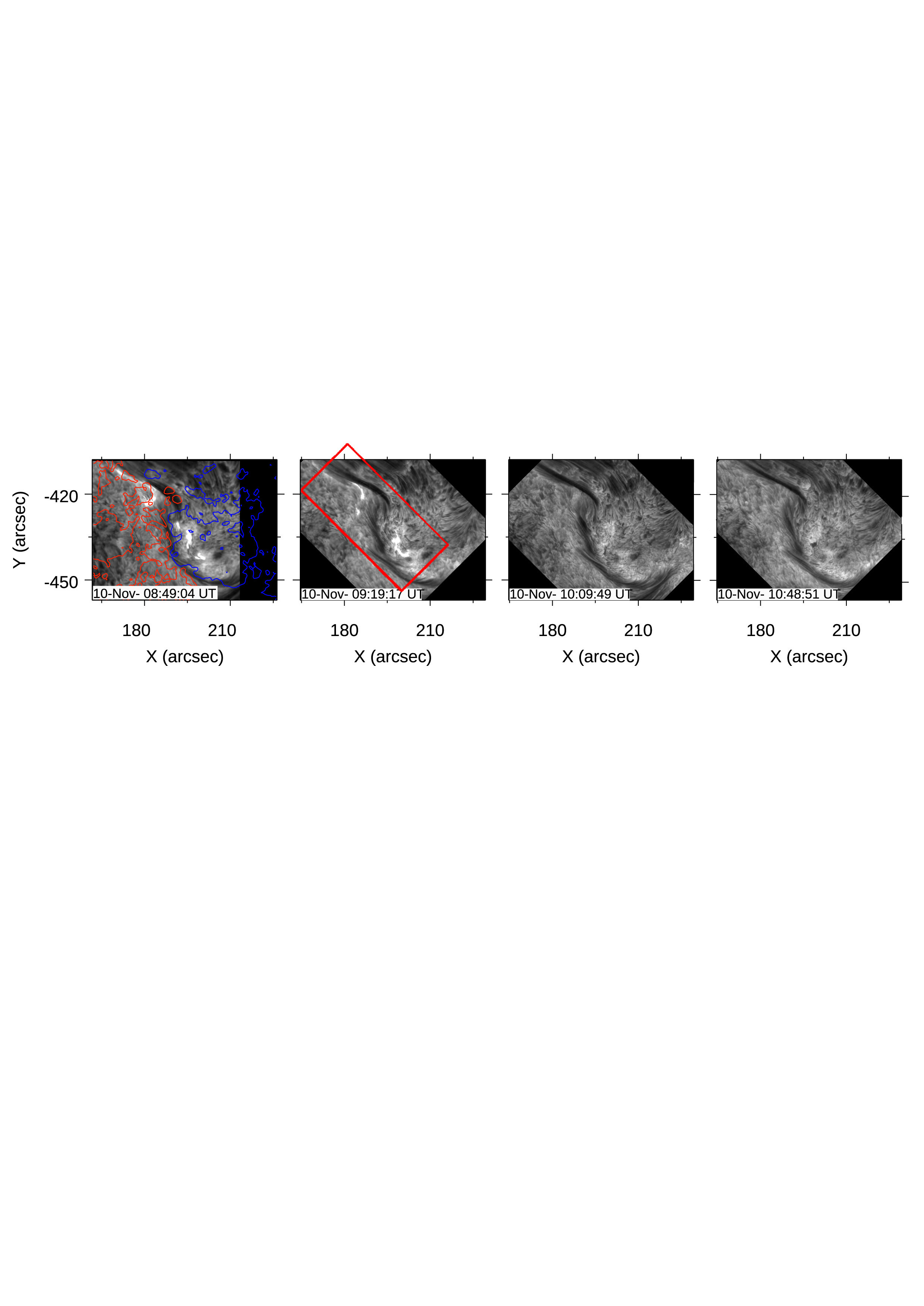}\\
\caption{Restored H$\alpha$ filtergrams acquired by the HiFI instrument at the GREGOR telescope showing the fine structure of the filament at various times on November 9 (top panels) and November 10 (bottom panels) November. The red and blue contours in first panels indicate the positive and negative magnetic flux density obtained from SDO/HMI at level equal to $\pm$200 G, respectively. The red box in the second bottom panel represents the GRIS acquired scanned FoV acquired between 09:17:32 and 09:37:05 UT. Movies of the HiFI data showing the filament in the time interval between 08:44 UT
and 11:02 UT (10:49 UT) on November 9 and 10 are available online.}
\label{fig4}
\end{figure*}

\subsection{H$\alpha$ filament fine structure}

Figure \ref{fig4} displays the restored H$\alpha$ filtegrams acquired by the HiFI instrument and shows the fine structure of the filament at various times on November 9 and 10 (top and bottom panels, respectively).
The filament is approximately 5\arcsec\ wide and longer than 50\arcsec\ (see the maps shown in Fig. \ref{fig4}). In particular, the first maps on November 9 and 10 (first column of Fig. \ref{fig4}) show the contours of locations of positive and negative magnetic fields from HMI data, marked in red and blue, respectively. On both days, the filament is located above the main PIL of the AR. The red box in the second bottom panel displays the area scanned by GRIS that covers the filament. A closer inspection of the top panels in Fig. \ref{fig4} for November 9 reveals a system of threads that is less compact on the northern side when compared to the southern one. This effect could be ascribed to the viewing angle of the structure. The filament appears progressively darker and more compact with time (see the online movies). Furthermore, on November 10 it is surrounded by active areas, which are seen as bright in the H$\alpha$ images. The filament appears to be composed of more than two systems. A new, darker, and more compact bundle emerges in the northern part. In particular, by comparing the IRIS \mg k$_{3}$, AIA 304 \AA\, and AIA 193 \AA\ maps (panels B, C and E in Fig. \ref{Fig1}) it is possible to note that this latter bundle, which becomes darker and more compact in H$\alpha$, does not seem to belong to the northern AFS, but rather resembles a continuation of the central filament. This interpretation is further supported by the magnetic configuration shown in panel F in Fig. \ref{Fig1}. From visual inspection, between November 9 to 10 (see Fig. \ref{fig4} from top panels to bottom panels and also the online movies) the thread orientation changes in a clockwise direction. However, this does not occur uniformly along the entire length of the filament. 
As generally expected at AFS footpoints, where velocities decrease (redshifted) and then accelerate to supersonic speeds, generating shocks and heating the surrounding atmosphere and higher layers, two homologous brightening events are detected at the edge of the northern AFS and on the western side of the filament at 08:49 UT and 09:19 UT on November 10 (see the first two bottom panels of Fig. \ref{fig4} and the online movie).

\begin{figure}
\includegraphics[scale=0.68,clip,trim=10 160 150 450]{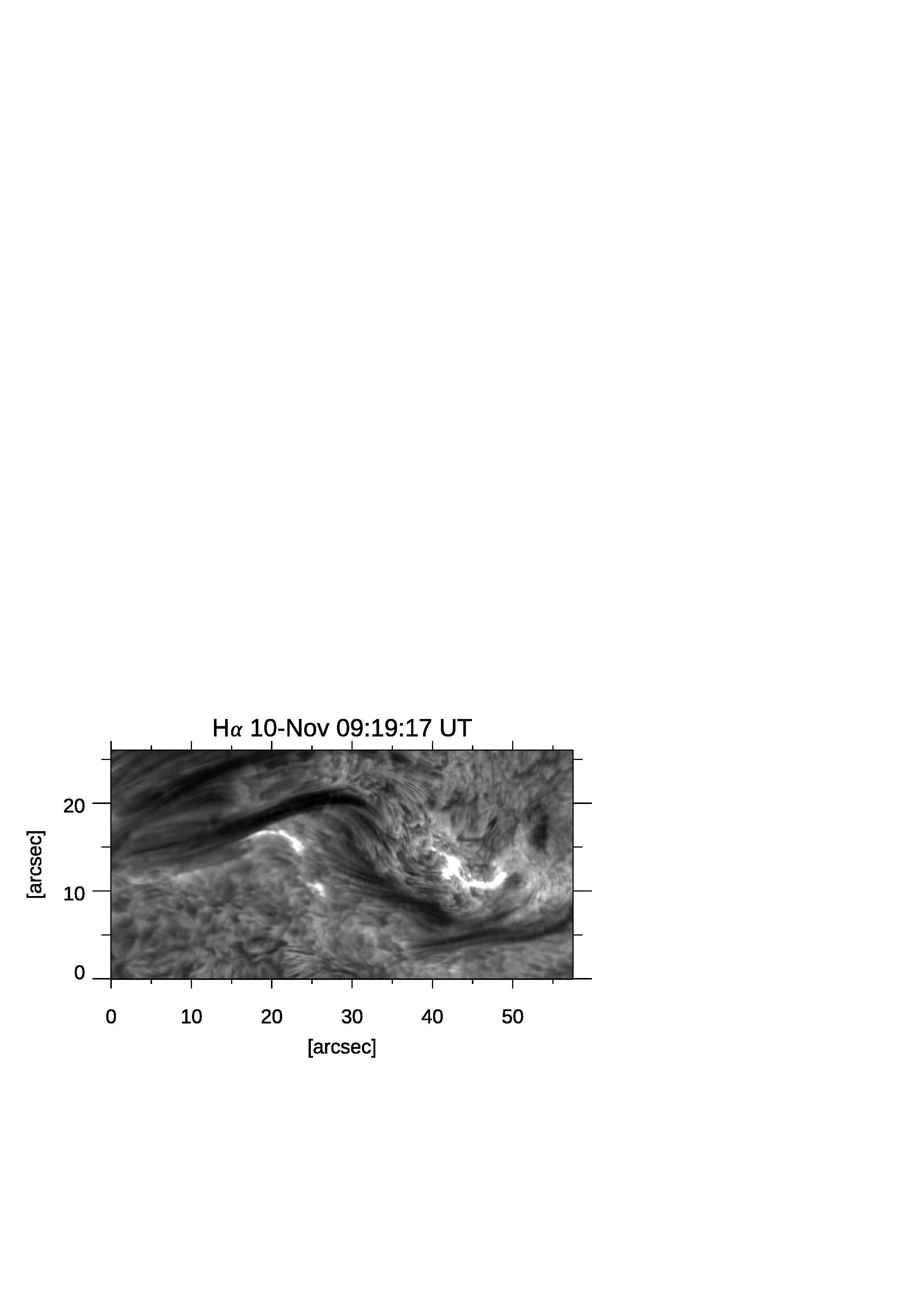}\newline
\includegraphics[scale=0.68,clip,trim=10 160 0 425]{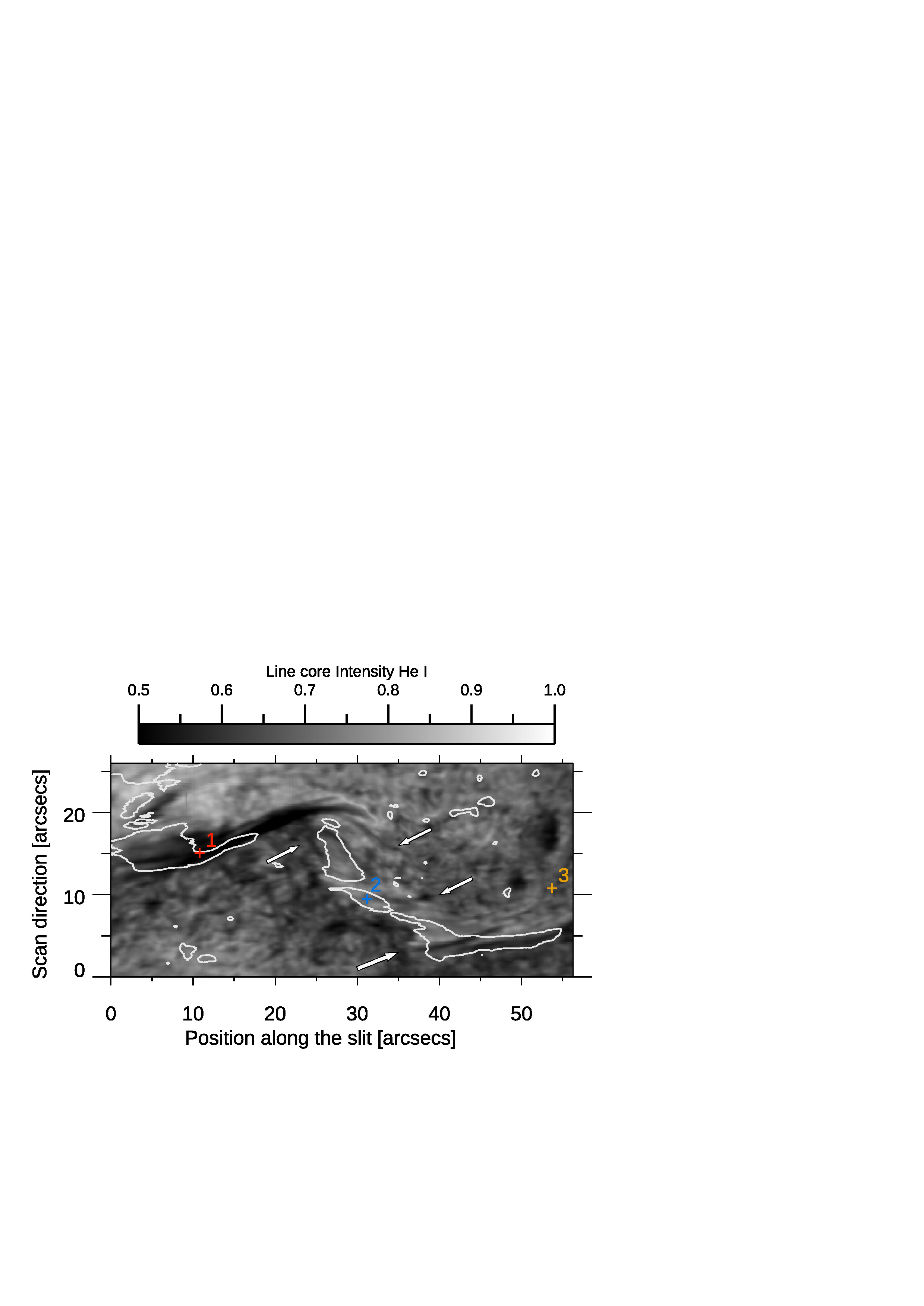}\\
\includegraphics[scale=0.2,clip,trim=70 0 0 40]{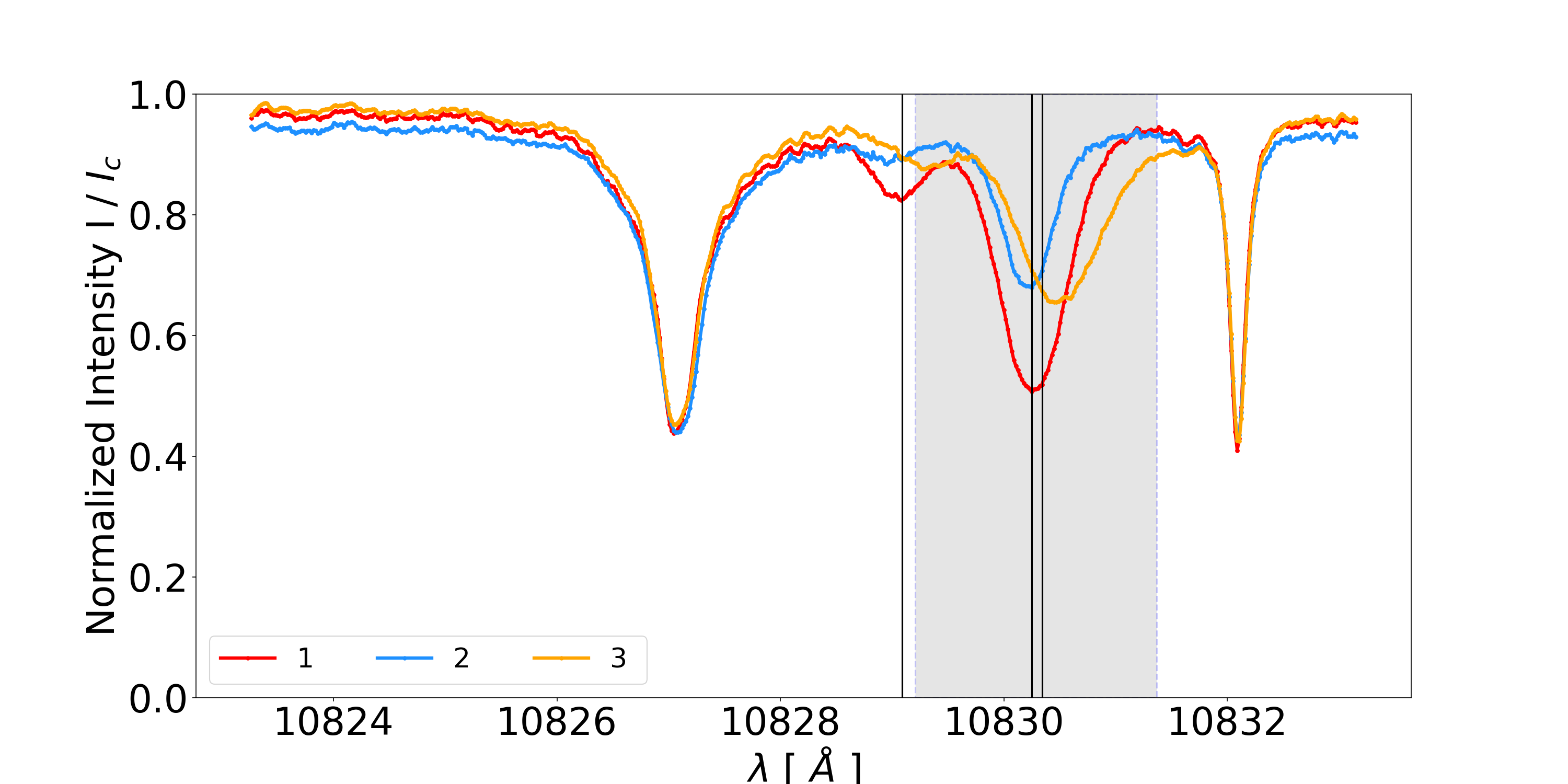}
\caption{Top panel: Zoom of the H$\alpha$ filtergram for reference. Middle panel: GRIS slit-reconstructed \ion{He}{i} line core intensity of the RoI acquired between 09:17:32 and 09:37:05 UT on November 10. Bottom panel: Stokes-I profiles, normalized to the continuum intensity, observed at the red, blue and orange (1, 2 and 3) positions shown in the \ion{He}{i} line core map. Arrows point to thin darker elongated threads where the optical thickness of the \ion{He}{i} line is higher than their own surroundings (Fig. \ref{fig4bis}). White contours represent plasma velocity of about $-1$ km s$^{-1}$. The shaded area indicates where the \ion{He}{i} line core is calculated (see for more detail Sect \ref{sec_gris}). The vertical lines represent the rest wavelengths for the three \ion{He}{i} components as reported in Table 1 of \citet{Kuckein2012b}. }
\label{fig4b}
\end{figure}

\subsection{\ion{He}{i} observation}

\begin{figure}
\includegraphics[scale=0.68,clip,trim=10 125 0 200]{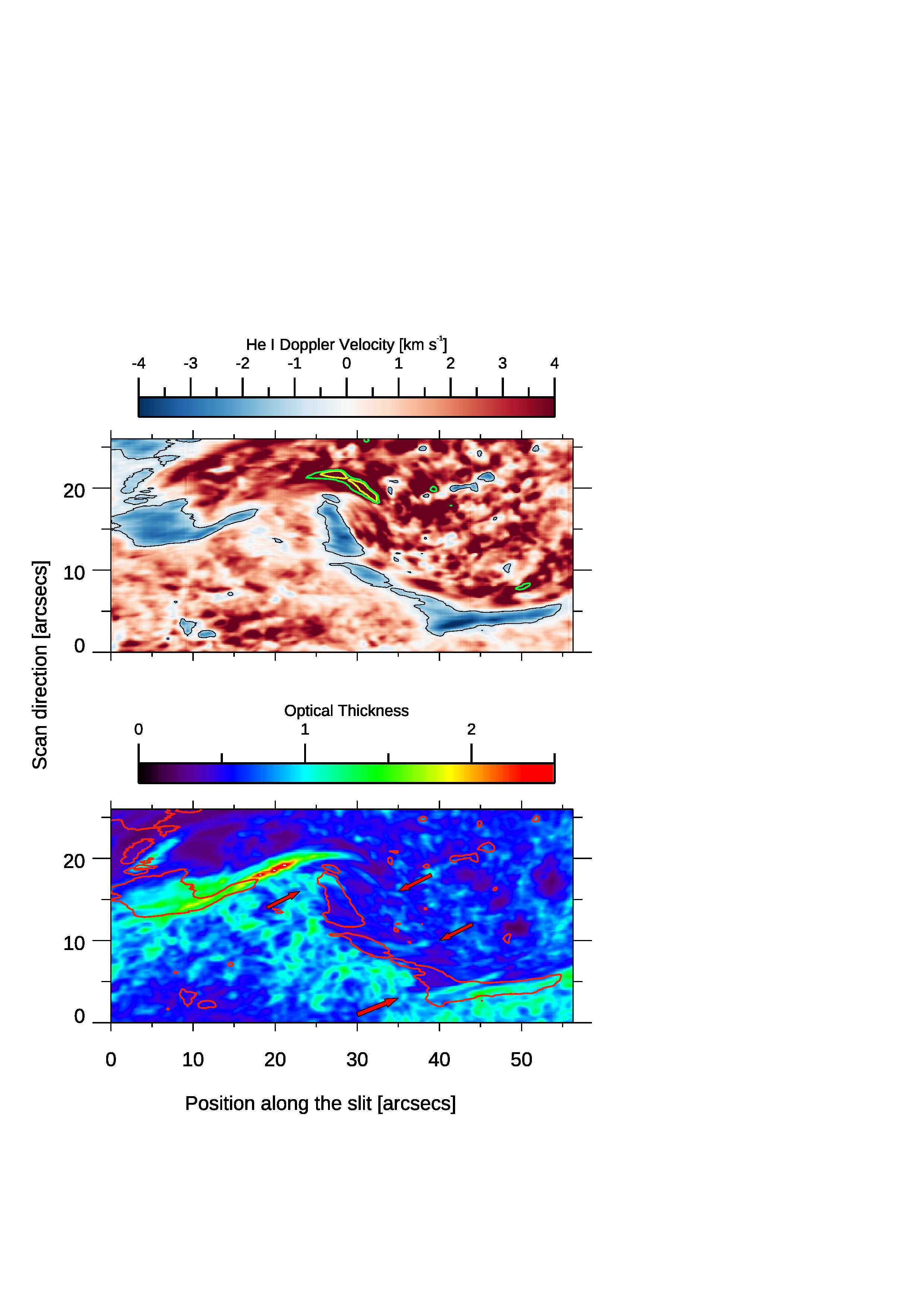}\\
\caption{\ion{He}{i} Doppler velocity and optical thickness of the GRIS observation, as calculated using the HAZEL code. Black and red contours represent plasma velocity of $-1$ km s$^{-1}$. Green and yellow contours indicate plasma velocity greater than 10 and 15 km s$^{-1}$. }
\label{fig4bis}
\end{figure}

Figure \ref{fig4b} display the slit-reconstructed line core intensity map of the \ion{He}{i} line (middle panel), and the corresponding H$\alpha$ image for a comparison (top panel). 
The \ion{He}{i} line-core intensity map shows a portion of the filament, outlined by the red box in  Fig. \ref{fig4}, as a thin, dark and elongated structure extending from 0 to 30\arcsec\ in x-direction. Additionally, there are a series of thin threads marked by the white arrows from 25\arcsec\ to 40\arcsec (x-direction). 
Interestingly, the filament does not appear as a dark, continuous structure along its full length in \ion{He}{i} 10830 \AA, unlike in H$\alpha$ observations. This may reflect variations in the PCTR across different segments of the filament, differences in the local EUV illumination, which we cannot directly measure, or a combination of both effects. These variations in \ion{He}{i} absorption may also correspond to the two filament segments discussed in the H$\alpha$ section, although this requires further verification. 
Red, blue, and orange crosses, labeled as 1, 2, and 3, are marked on the \ion{He}{i} line core intensity map to indicate specific pixels belonging to the QS (label 3), and to dark and bright regions within the filament (labels 1 and 2, respectively). The spectral profiles for each of these positions, within the \ion{He}{i} spectral window, are shown in the bottom panel of Fig.~\ref{fig4b}.
The normalized spectrum shows that the absorption profile in the central bright part of the filament (blue profile labeled "2" in the middle panel) is significantly shallower, with an intensity reduction of about 30\%, compared to the darkest region (red profile). This shallower profile is more similar to that observed in the QS (orange profile). 
This peculiarity could be ascribed to specific mechanisms of \ion{He}{i} triplet line formation, which is influenced by both EUV illumination from the surrounding corona and the presence of the filament PCTR, as discussed by \citet{Andretta1997,Labrosse2002,Labrosse2004}.  
Moreover, thin elongated, darker threads are visible in the central portion of the filament in the \ion{He}{i} line core intensity map (see white arrows in Fig. \ref{fig4}).

Maps of Doppler velocity and the slab optical thickness \footnote{For an explanation of how this parameter is calculated, see \url{https://aasensio.github.io/hazel/equations.html}.} of \ion{He}{i} as retrieved by the HAZEL2 code are shown in Fig. \ref{fig4bis}. The Doppler velocity map mainly displays two blue-shifted areas in correspondence of the AR filament while redshifted plasma is observed almost elsewhere. The darker and elongated part of the filament, as seen in the \ion{He}{i} line core, corresponds to the new structure that previously appeared in H$\alpha$. This region displays both upward (between X coordinates from 0\arcsec\ to 15\arcsec) and downward motions (between X coordinates from 10\arcsec\ to 30\arcsec). In particular, the observed redshifted component reaches values exceeding 10 km s$^{-1}$ (as indicated by the green and yellow contours, corresponding to velocities of 10 and 15 km s$^{-1}$, respectively). This redshift may be associated with the footpoints of the northern AFS, and possibly also with those of the filament, suggesting a partial spatial overlap between the two structures in the chromosphere. 
The region characterized by a decrease in the \ion{He}{i} absorption profile (see the blue profile in Fig \ref{fig4} bottom panel) displays upward motions reaching velocities up to $-3.5$ km s$^{-1}$, with an average velocity of $-1.9$ km s$^{-1}$ (within the black contour in Fig. \ref{fig4bis}). This area also exhibits a relatively low optical thickness, below 1 (see bottom panel of Fig. \ref{fig4bis}).  
Notably, darker, thin threads, indicated by the red (or white) arrows, are visible in the region and are associated with slightly higher optical thickness values, around 1, compared to their surroundings.
In contrast, the darker, elongated structure, has an optical thickness values ranging from 1 to 2.5, consistent with those reported by \citet{Baso19}.

\begin{figure*}
\centering
\includegraphics[scale=0.52,clip,trim=10 295 260 
200]{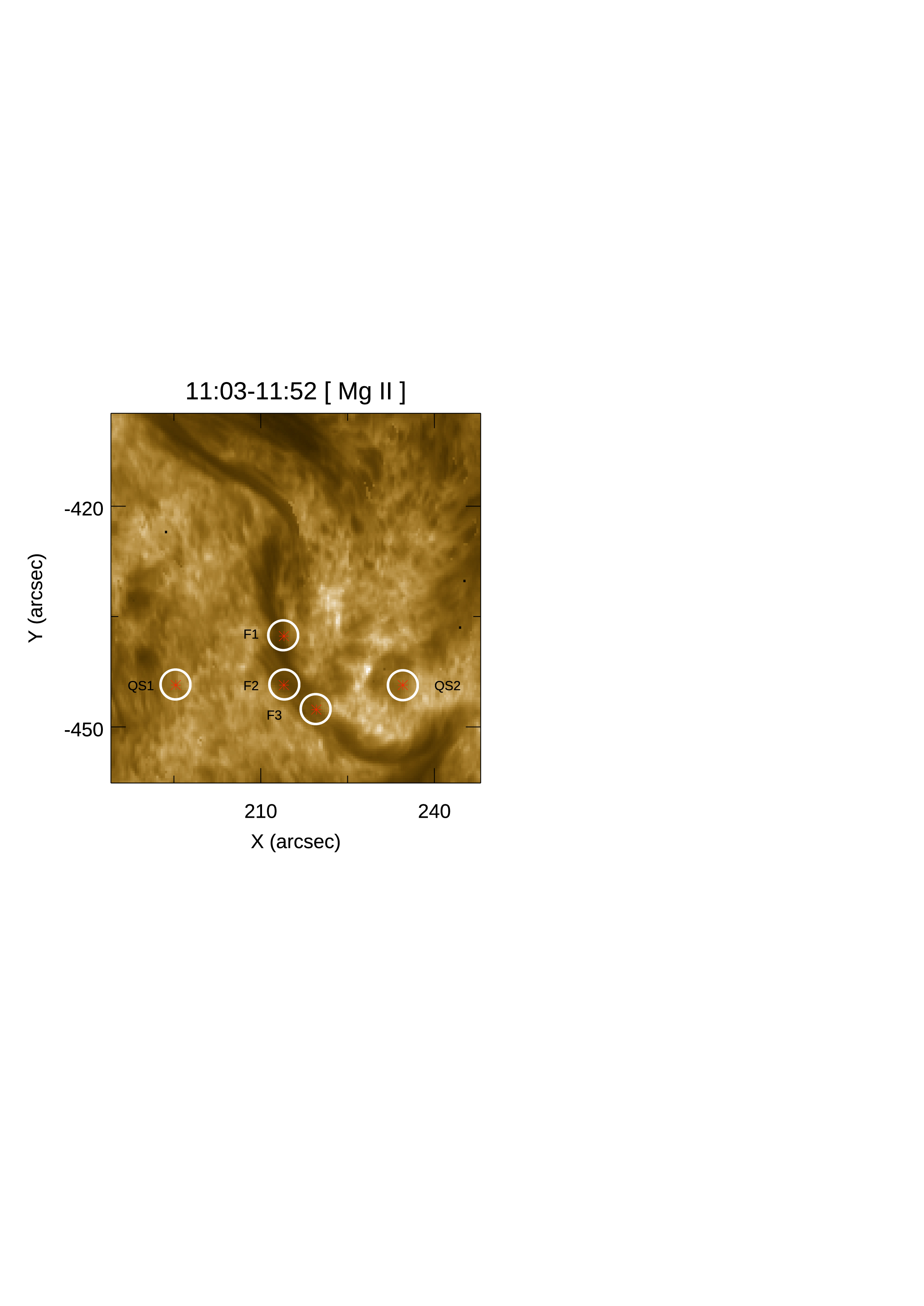}
\includegraphics[scale=0.52,clip,trim=10 295 260 
200]{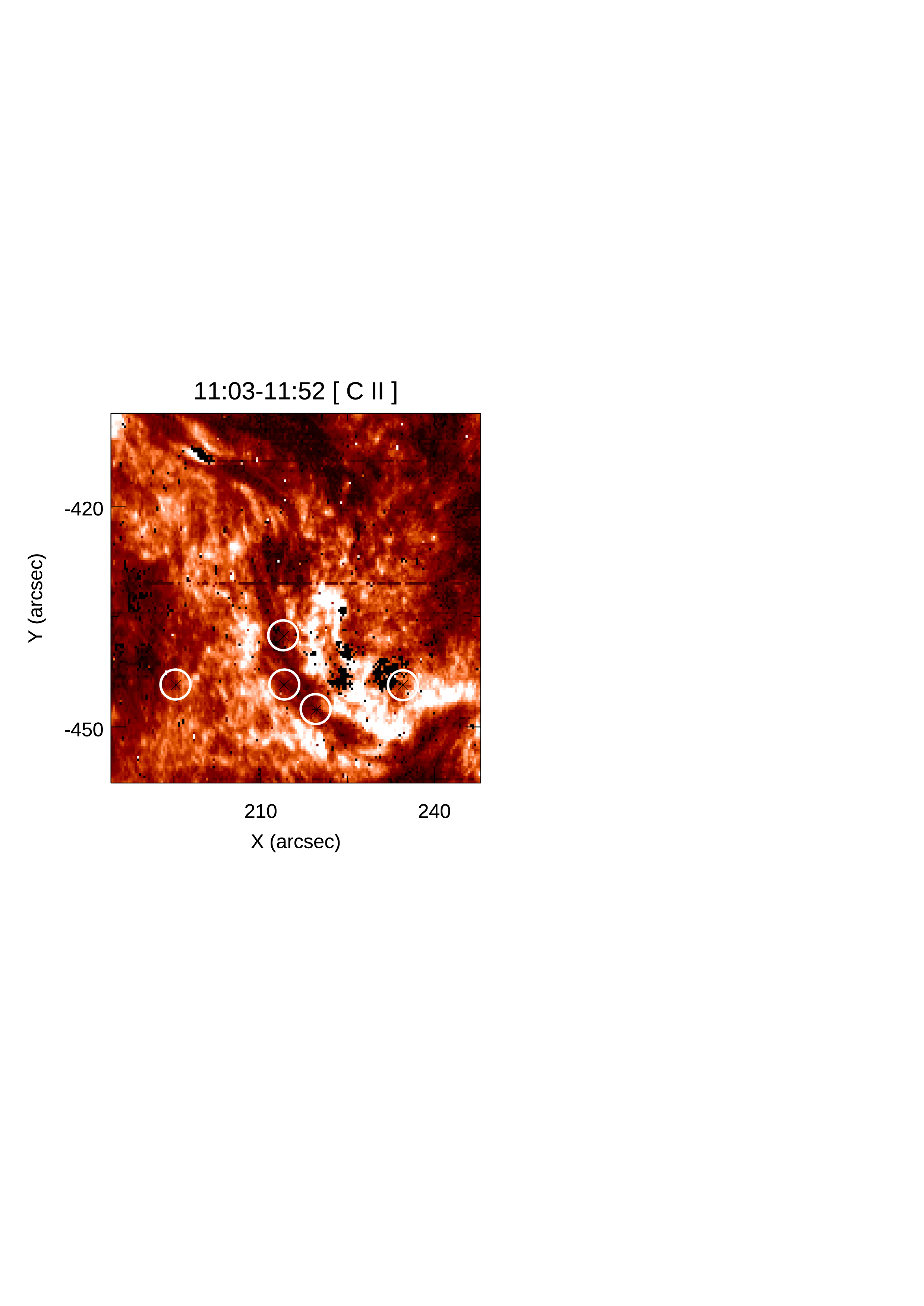}
\includegraphics[scale=0.52,clip,trim=10 295 260 
200]{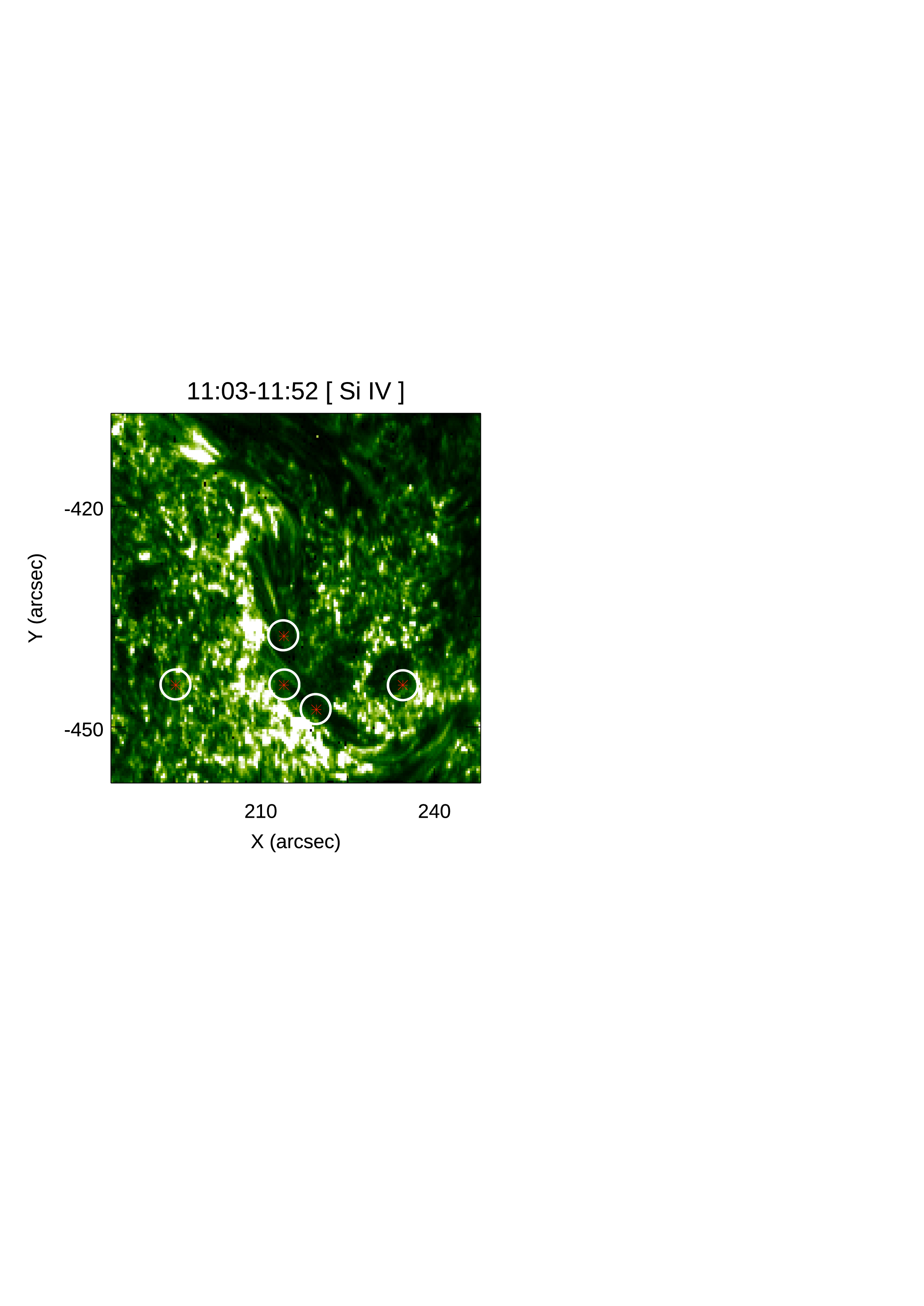}\\
\includegraphics[scale=0.48,clip,trim=90 130 120 
50]{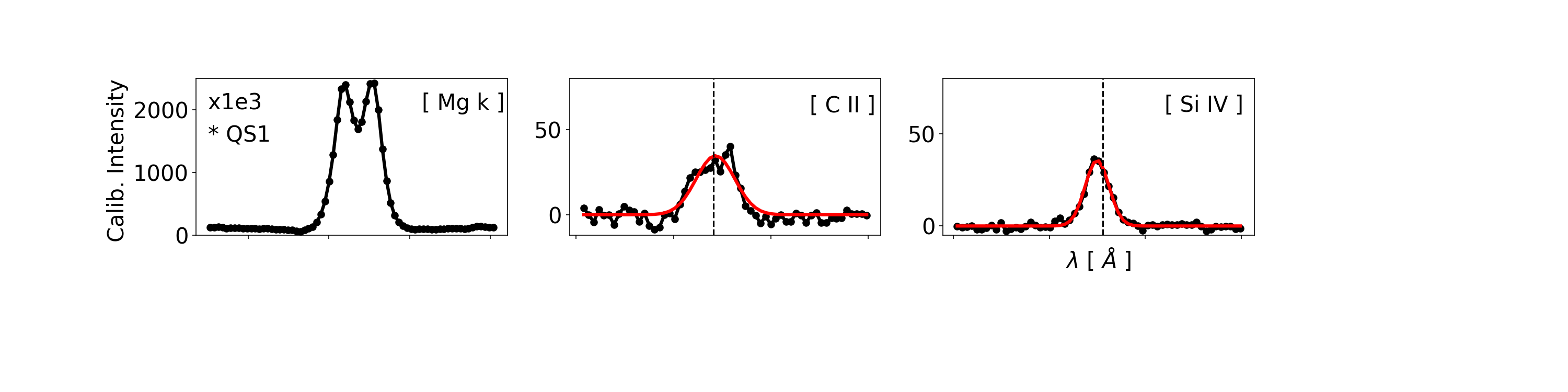}
\includegraphics[scale=0.48,clip,trim=90 130 120 
50]{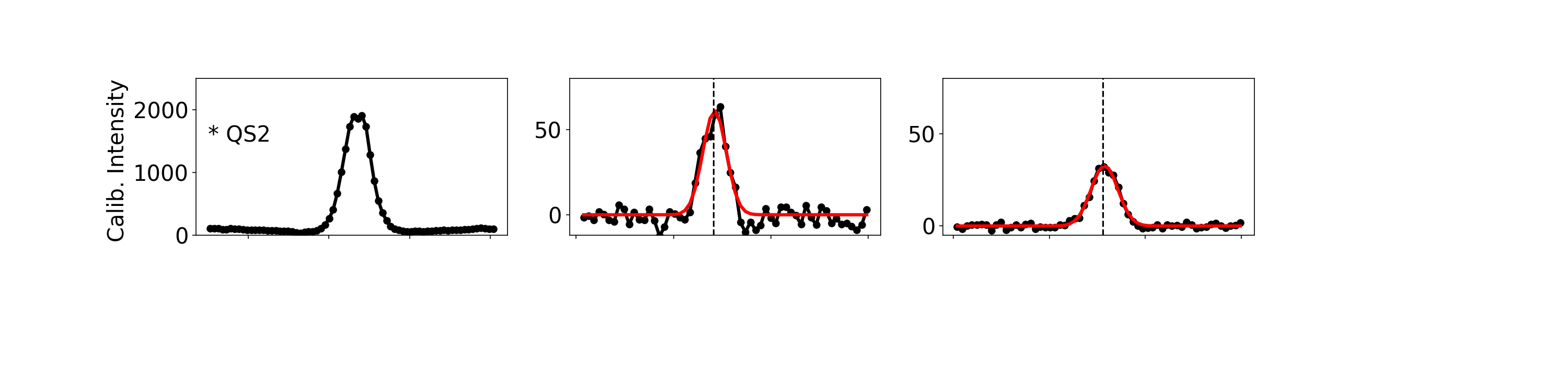}
\includegraphics[scale=0.48,clip,trim=90 130 120 
50]{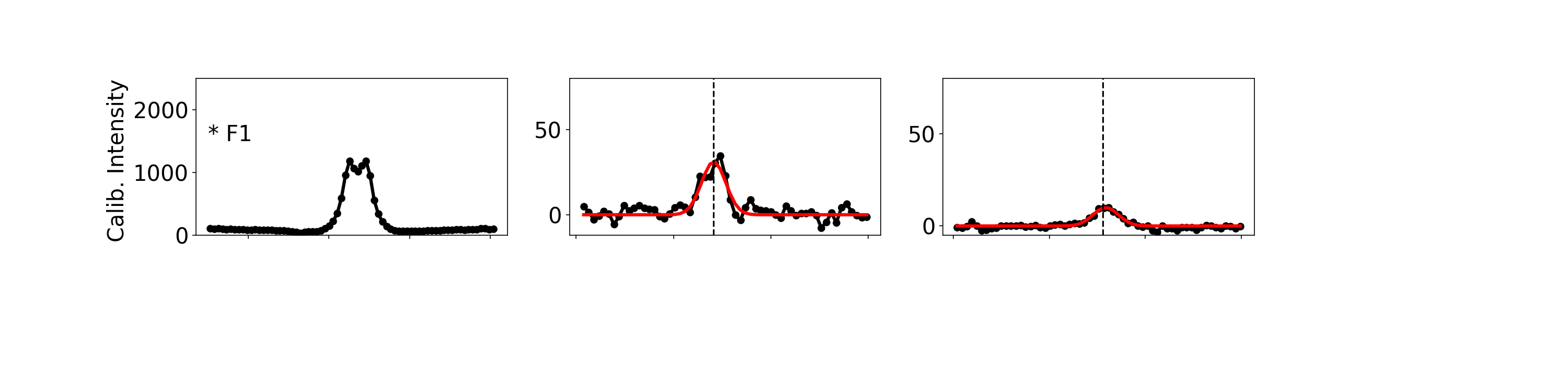}
\includegraphics[scale=0.48,clip,trim=90 130 120 
50]{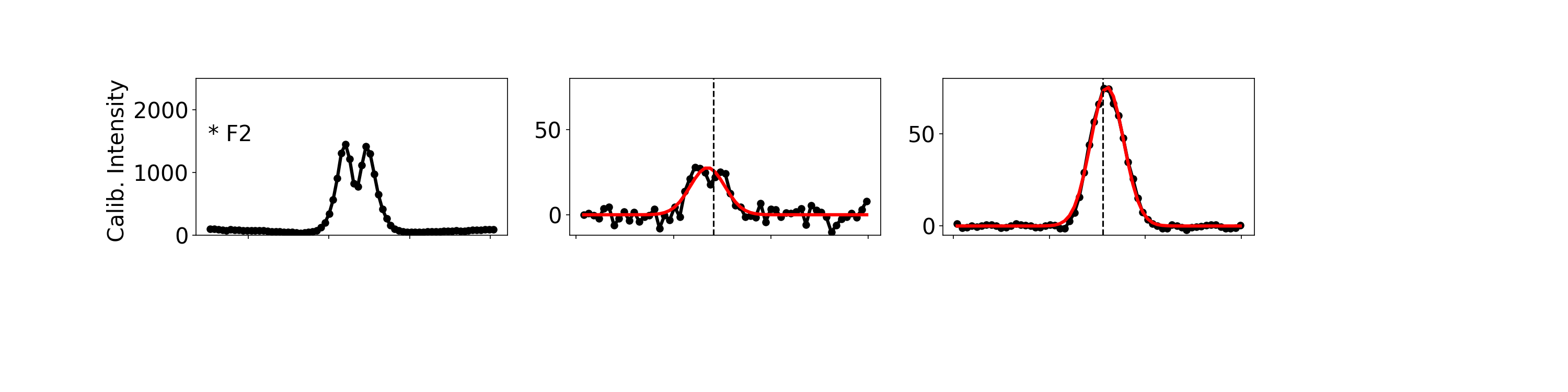}
\includegraphics[scale=0.48,clip,trim=90 90 120 
50]{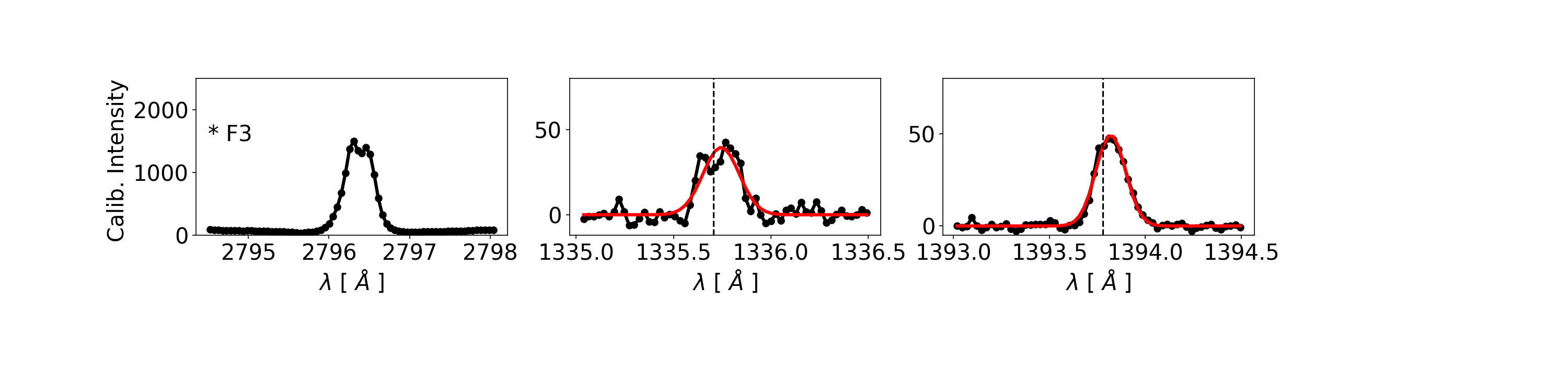}
\caption{From top to bottom: Peak intensity maps and spectral profiles in the locations marked with an asterisk in the respective intensity maps and their relative fits of \mgk, \si 1394 \AA\ and \ion{C}{ii} 1335 \AA\ during the middle IRIS raster scan (11:03 - 11:52 UT). Original spectral line profiles (in black) and corresponding Gaussian fits in red color (for \si and \ion{C}{ii} spectral lines).}
\label{cal1}
\end{figure*}

\begin{figure}
\includegraphics[scale=0.5,trim=20 417 200 180]{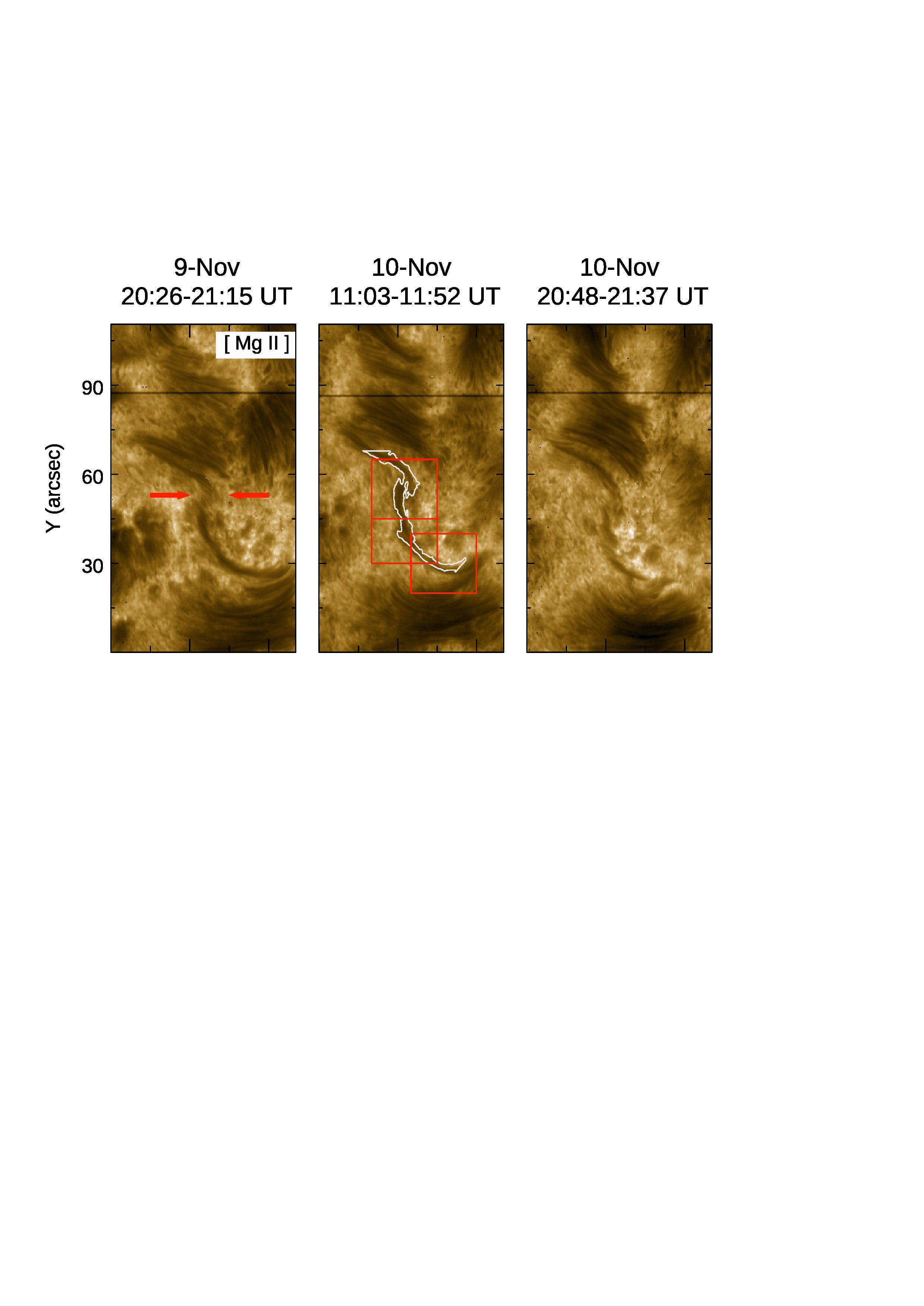}\\
\includegraphics[scale=0.5, clip, trim=20 417 0 200]{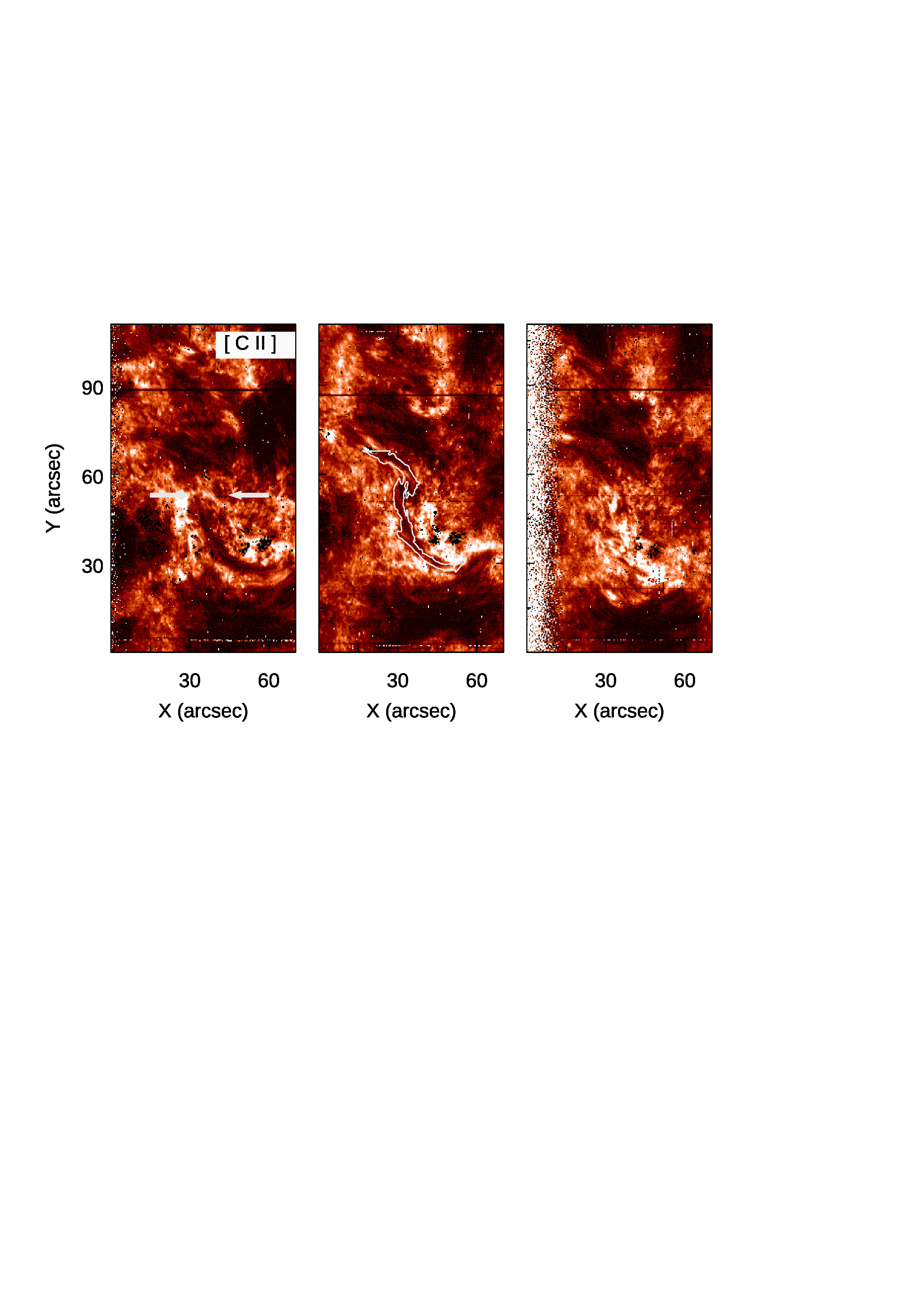}\\
\includegraphics[scale=0.5, clip, trim=20 375 0 200]{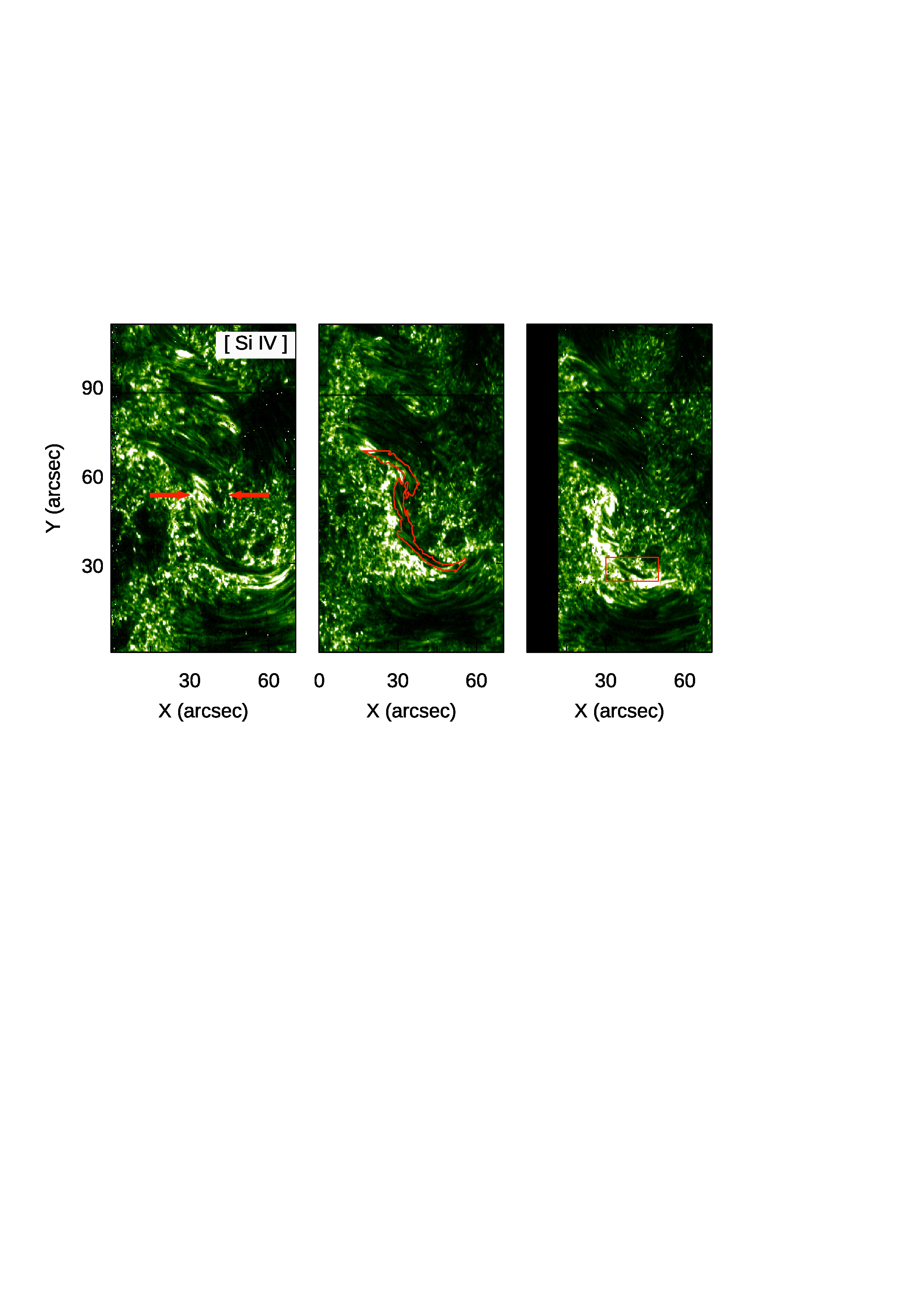}\\
\caption{Peak Radiance maps \mg (top), \ion{C}{ii} (middle) and \si (bottom). These maps result from the three IRIS rasters reported in Table \ref{table1}. Contours in all middle panels indicate the high-resolution GRIS H$\alpha$ filament contour. The red boxes represent the three panels shown in Fig.\ref{Fig7b}. }
\label{Fig5}
\end{figure}

\subsection{UV IRIS observations}

Figure \ref{cal1} shows the IRIS peak intensity maps within the RoI, acquired during the raster scan on November 10, between 11:03 UT and 11:52 UT. The maps correspond to the three brightest IRIS UV lines, ordered by increasing formation temperature: \mg k$_{3}$, \ion{C}{ii} 1336 \AA, and \si 1393 \AA. Spectral profiles at five representative pixel positions, marked with asterisks on the maps, are also shown. These pixels were selected to represent different spectral behaviors observed across the FoV. For both the \ion{C}{ii} and \si pixel profiles, single-Gaussian fits to the observed profiles are shown in red.

Within the filament, the \mgk\ line profiles (e.g., at F1, F2, and F3) exhibit typical spectral features: two emission peaks (k${2}$) separated by a central absorption dip (k${3}$). In contrast, outside the filament, such as at the QS2 location, which belongs to a magnetized area due to the presence of a nearby pore and sunspot (see Fig. \ref{Fig1} G), the profiles may lack the k${2}$ peaks and appear fully in emission. While these differences do not affect the measurements of the k$_{3}$ Doppler shift, they can influence other spectral diagnostics (see Fig. \ref{Fig7} and related discussion). As shown by \citet{Rathore2015a}, \ion{C}{ii} line profiles may exhibit either a single- or a double-peaked profile, depending on the variation of the source function and the presence of LoS velocity gradients. A central absorption dip forms when the source function decreases before the optical depth reaches unity. Velocity gradients can Doppler-shift the absorption profile, distorting symmetric double peaks into asymmetric single peaks. For a comprehensive discussion of \ion{C}{ii} line formation, refer to \citet{Rathore2015a,Rathore2015b,Avrett2013}. In our data, QS profiles are not uniformly single-peaked: for example, the QS2 profile is single-peaked, while the QS1 profile shows poorly defined peaks. In contrast, the filament profiles (F1, F2, and F3) consistently display double-peaked \ion{C}{ii} profiles. 

The last column of Figure \ref{cal1} presents the corresponding \si profiles at the five selected pixel positions. In all filament positions, the signal is well above the noise level, ensuring the robustness of the Gaussian fits and, consequently, the accuracy of the Doppler-shift estimates derived from the line centroid, given that the spectral profiles exhibit a clear Gaussian shape.

Figure \ref{Fig5} reports peak intensity maps for the \mg k, \ion{C}{ii} and \si spectral lines for the entire FoV observed, respectively. Although the first IRIS raster scan was acquired approximately ten hours after the GREGOR high-resolution H$\alpha$ observations, a comparison reveals that the morphology of the filament in the first \mgk scan (acquired between 20:26 and 21:15 UT on November 9) is similar to that seen in the GREGOR H$\alpha$ images (see bottom panels of Fig. \ref{fig4}) taken on November 10, 08:00 UT. Due to the spatial resolution of IRIS and the different formation heights of the spectral lines, the fine structure of the filament seen in the ground-based H$\alpha$ observations is not resolved in the \mgk peak intensity maps. Individual threads along the filament axis are distinguishable in the \mgk peak intensity. Nonetheless, a filamentary but less compact structure, somewhat resembling that seen in H$\alpha$, can still be identified in the region X=[30\arcsec,45\arcsec] and Y=[45\arcsec,60\arcsec], as indicated by the two red arrows in the top left panel. This gap region shows localized brightening activity in both the \ion{C}{ii} and \si 1393 \AA\ intensity maps. This latter (bottom-left panel) also displays the presence of filamentary threads at the western edge of the gap, as well as within the central brightening area. Given the observable temporal changes, these features may indicate ongoing localized reconnection, where threads are changing from inclined to vertical configurations, consistent with the findings of \citet{Zhou2016}.

The second IRIS scan was acquired just at the end of the H$\alpha$ ground-based observations. In this case, the \mgk and H$\alpha$ images appear similar, both clearly showing barbs and fine threads along the filament. After about nine hours, during the third IRIS scan (taken between 20:48 and 21:37 UT; see top right panel) the filament structure seems to undergo structural changes. Rather than a continuous structure, the filament now appear as a bundle of filamentary threads, with a brighter area on its western side. 

Even though the \ion{C}{ii} and \si maps are noisier than those in the \mgk line, the filament remains well recognizable in the first two raster scans (first two middle and bottom panels of Fig. \ref{Fig5}). However, at the eastern and western edges of the filament, the peak intensity of the \ion{C}{ii} ~1336 \AA\ can not be reliably measured due to lower signal levels and the presence of more complex line profiles. The morphology of the filament on November 10 resembles that seen in the H$\alpha$ observations, both in terms of length and dimension (see the white contour in the middle panel of Fig. \ref{Fig5}). Brightening areas are visible at the edge of the filament and the last scan (right middle panel of Fig. \ref{Fig5}) clearly displays small-scale brightening events occurring within the filament.

In contrast with previously reported SUMER observations in \si lines \citep{Kucera1999},  the filament is clearly detectable in the \si peak intensity during the first and second raster scans. Similar to the \ion{C}{ii} maps, a localized bright region is observed surrounding the filament on its western side. The filament is still detectable under the H$\alpha$ red contour at 11:02-11:52 UT on November 10. By the time of the final raster scan, this bright region has expanded, appearing to cover much of the filament. Furthermore, several elongated bright patches are observed between the threads, while only a small portion of darker filamentary structure is visible (see the red box in the right bottom panel of Fig. \ref{Fig5}). These small-scale brightening events do not appear to be fully co-spatial with those seen in the \ion{C}{ii} data.

\begin{figure*}
\centering
\includegraphics[scale=0.8,clip,trim=20 417 0 180]{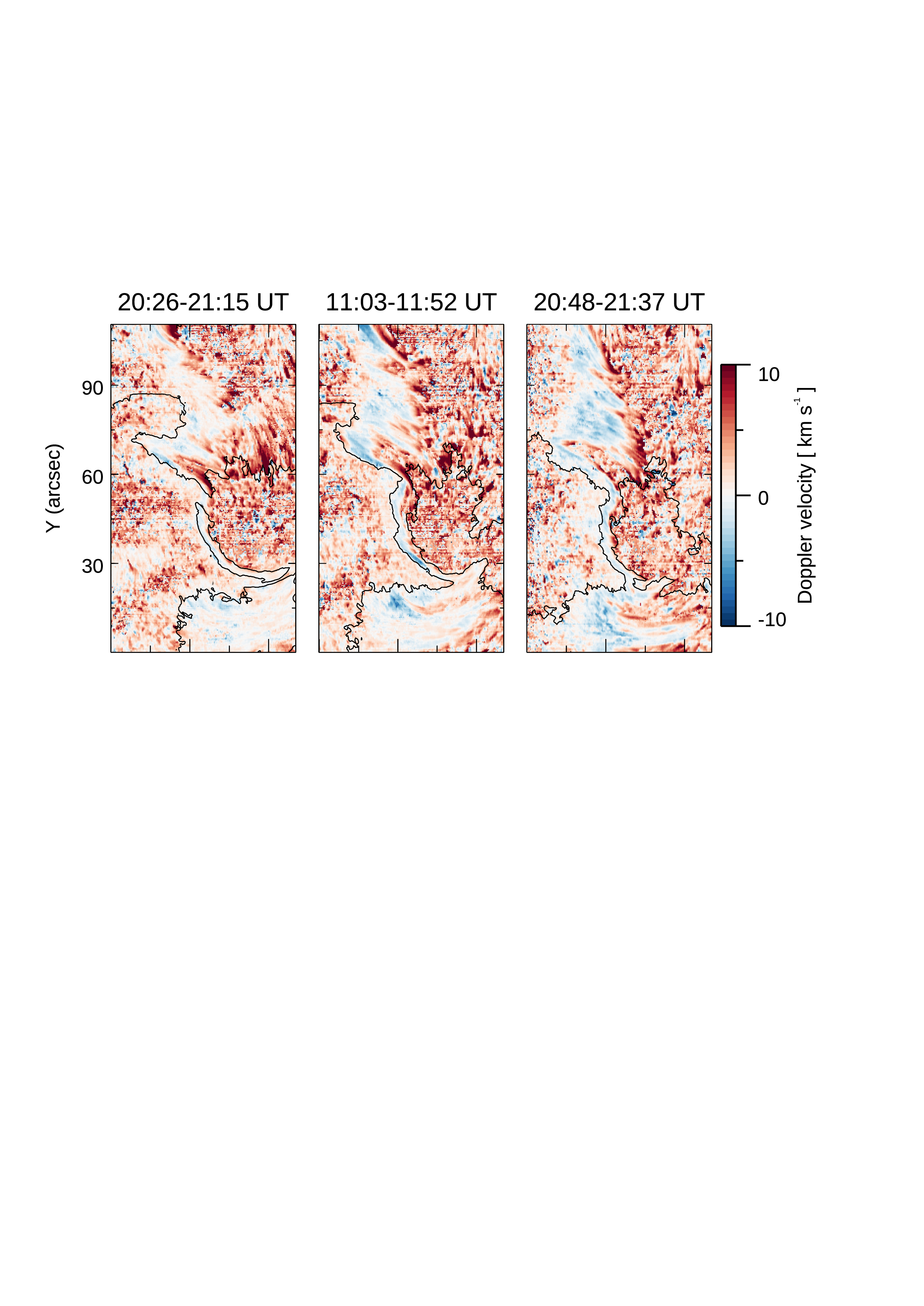}\\
\includegraphics[scale=0.8, clip, trim=20 417 0 200]{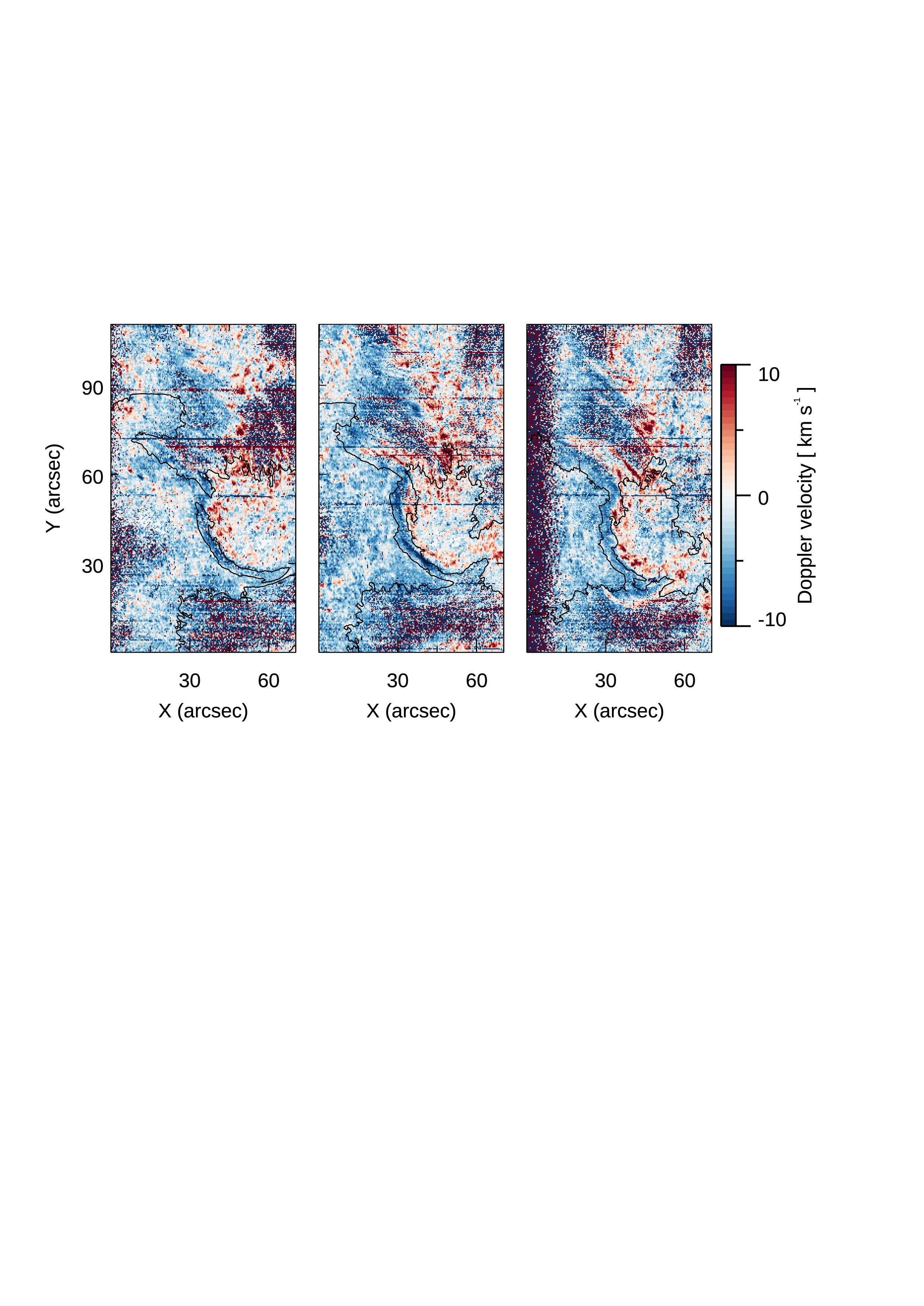}\\
\includegraphics[scale=0.8, clip,trim=20 370 0 200]{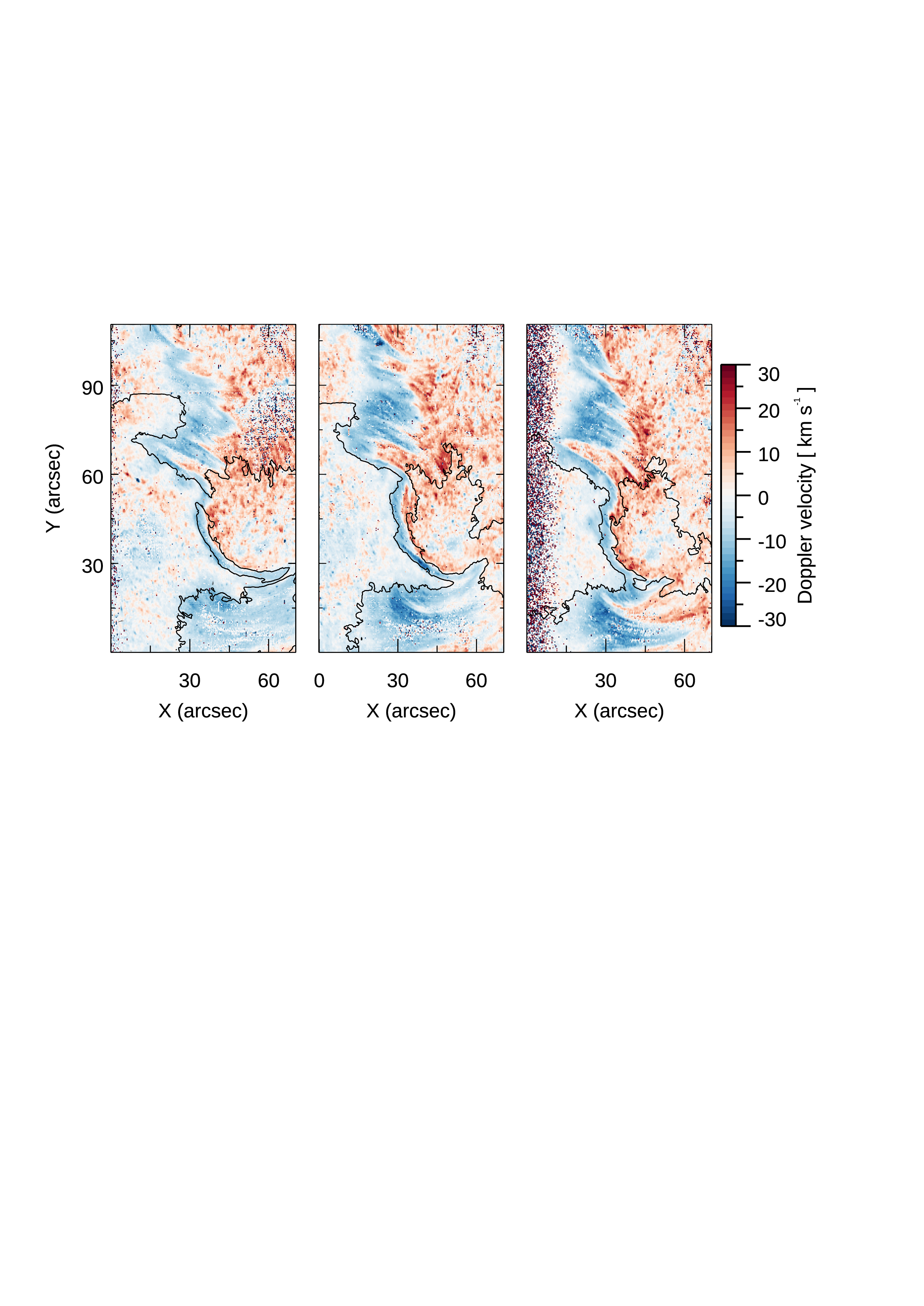}\\
\caption{Velocity maps for \mg (top), \ion{C}{ii}  (middle) and \si (bottom) observations. In all panels, black contour represents \mg k$_{3}$ intensity.}
\label{Fig6}
\end{figure*}

\begin{figure*}
\includegraphics[scale=0.85, clip, trim=0 448 0 260]{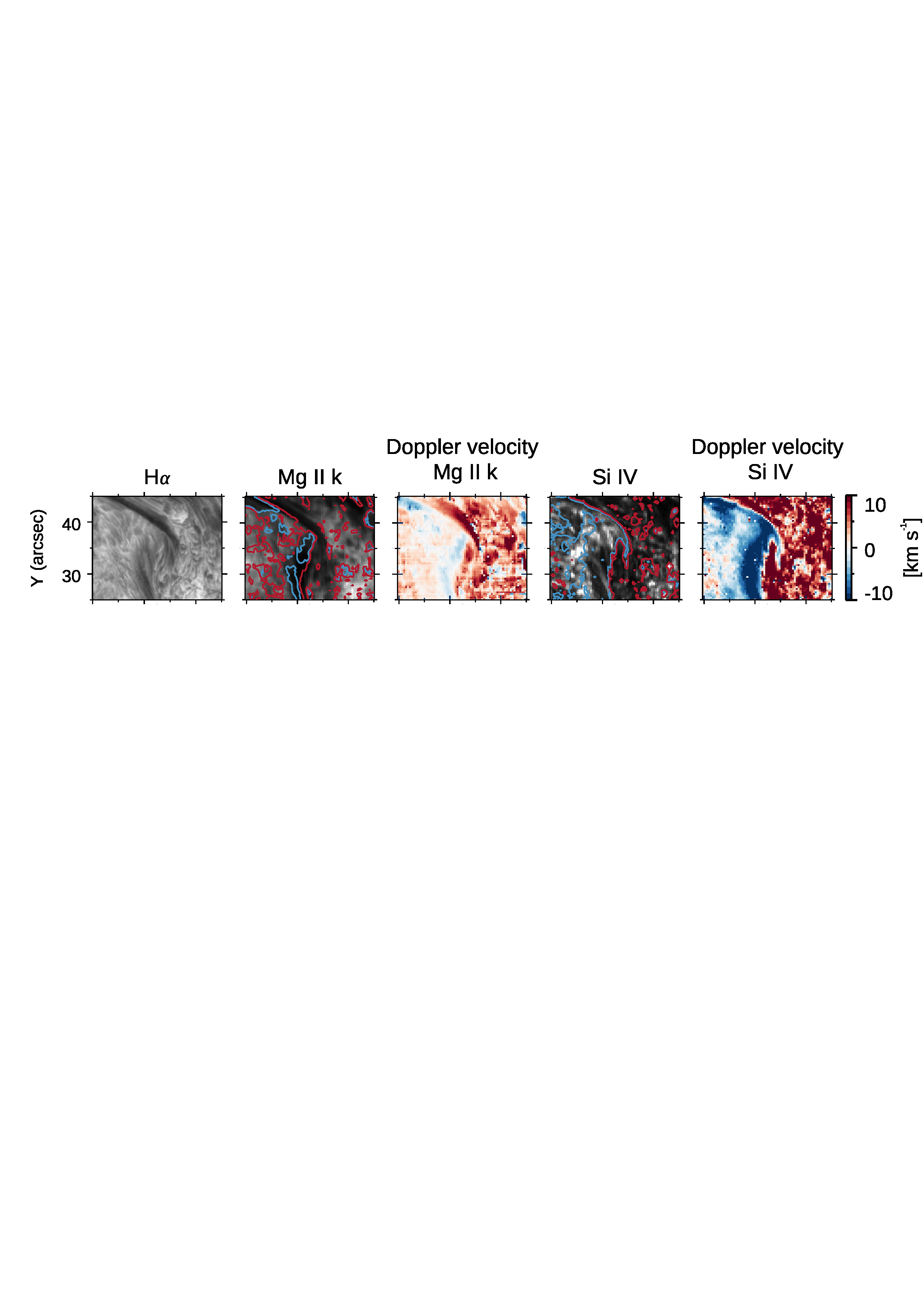}\\
\includegraphics[scale=0.85, clip, trim=0 448 0 300]{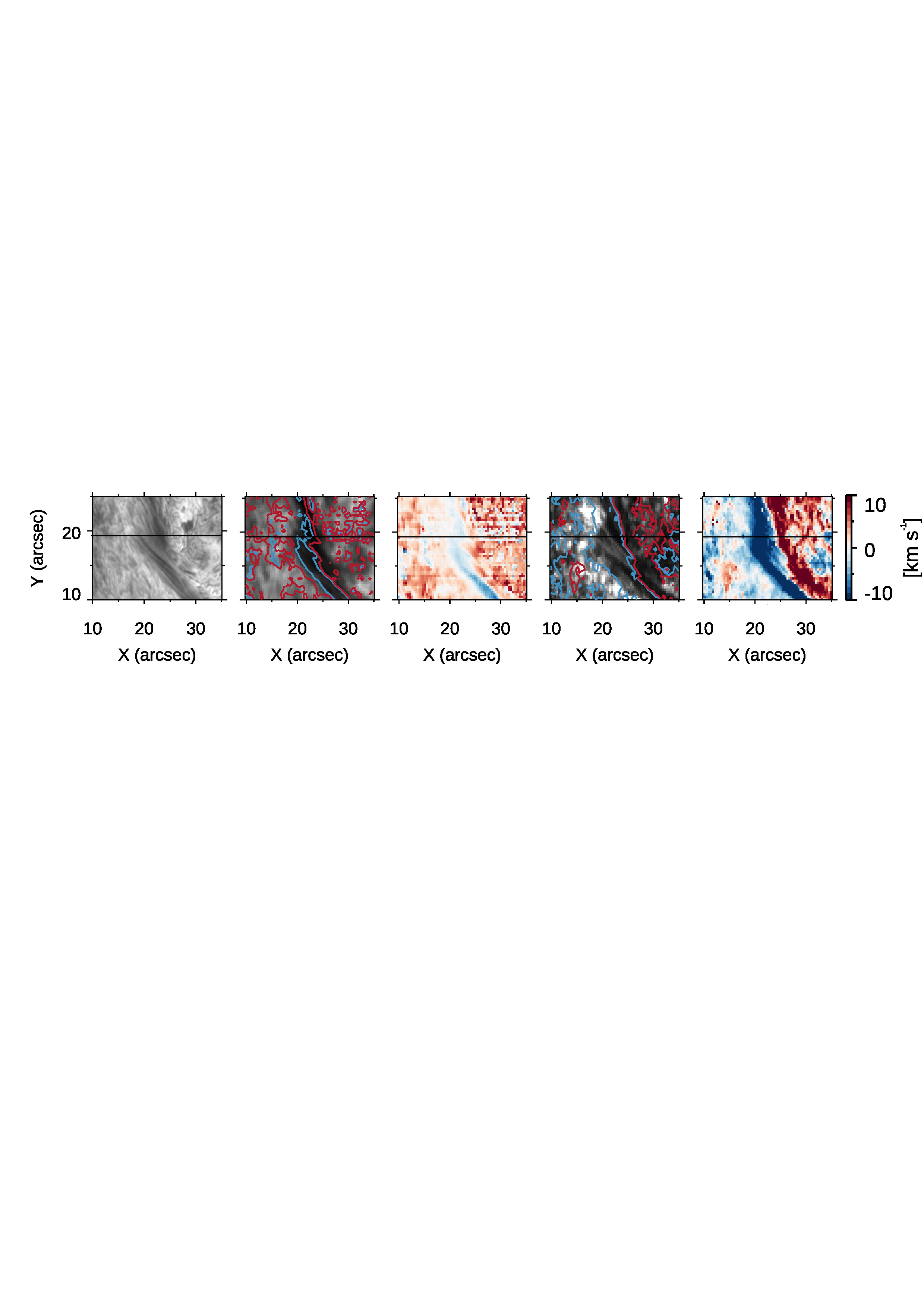}\\
\includegraphics[scale=0.85, clip, trim=0 410 0 300]{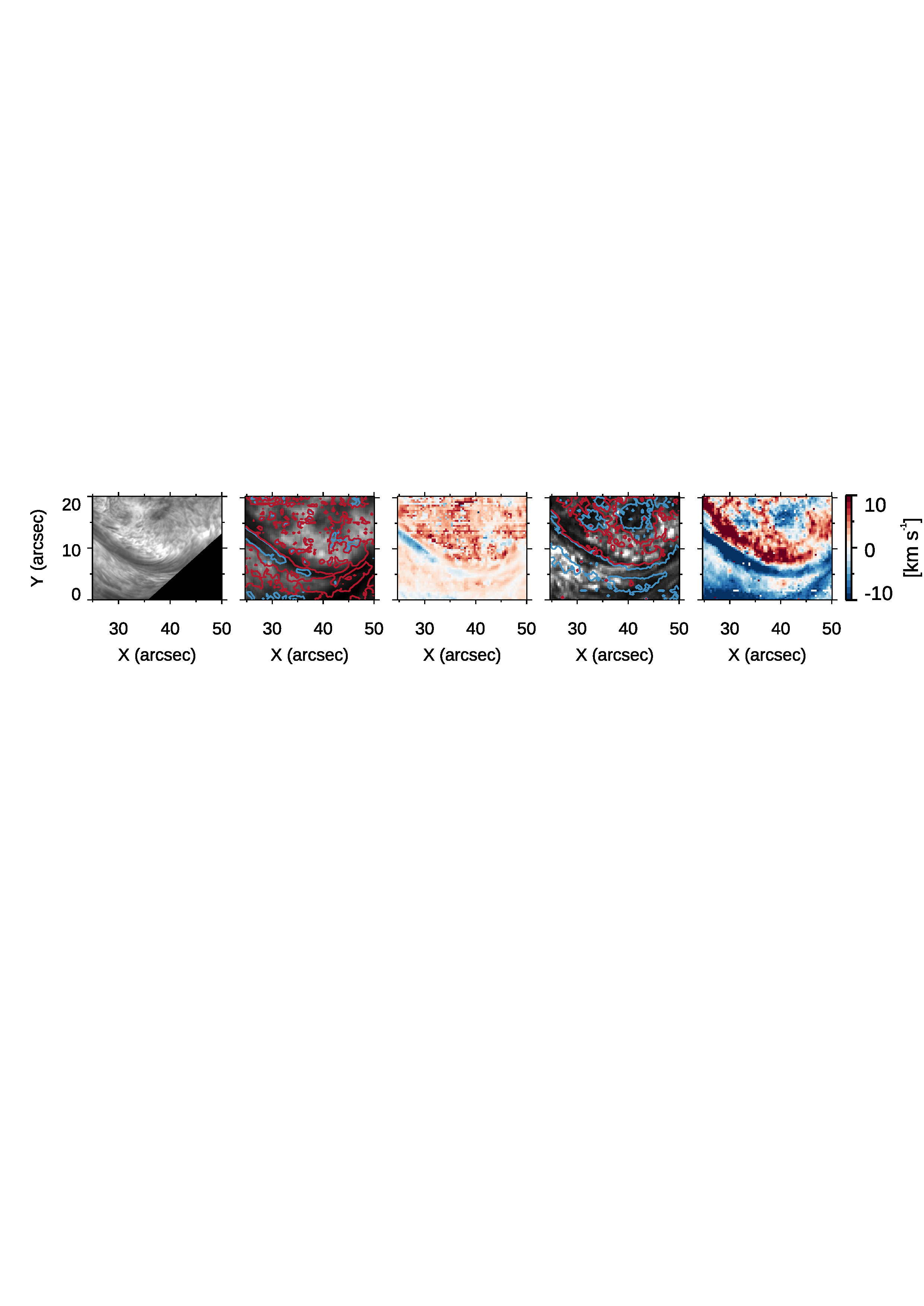}
\caption{Comparison of H$\alpha$ and UV \mg and \si IRIS intensity and Doppler velocity. From left to right: H$\alpha$ intensity, \mg k$_{3}$ and Doppler velocity, \si peak intensity and \si Doppler velocity maps for three RoIs  (see the red boxes in the k$_{3}$ intensity middle panels of Fig. \ref{Fig5}). The \mg k$_{3}$ and \si Doppler velocity maps are shown in the range of [-10, 10] and [-30,30] km s$^{-1}$ as the original maps of Fig \ref{Fig7}. In the \mg k$_{3}$ and \si intensity and Doppler velocity maps, the red and blue contours represent upflows (–1 and –3 km s$^{-1}$) and downflows (1 and 3 km s$^{-1}$), respectively, corresponding to the Doppler velocities of \mg k$_{3}$ and \si. }
\label{Fig7b}
\end{figure*}

\begin{figure*}
\includegraphics[scale=0.2,trim=30 30 60 20]{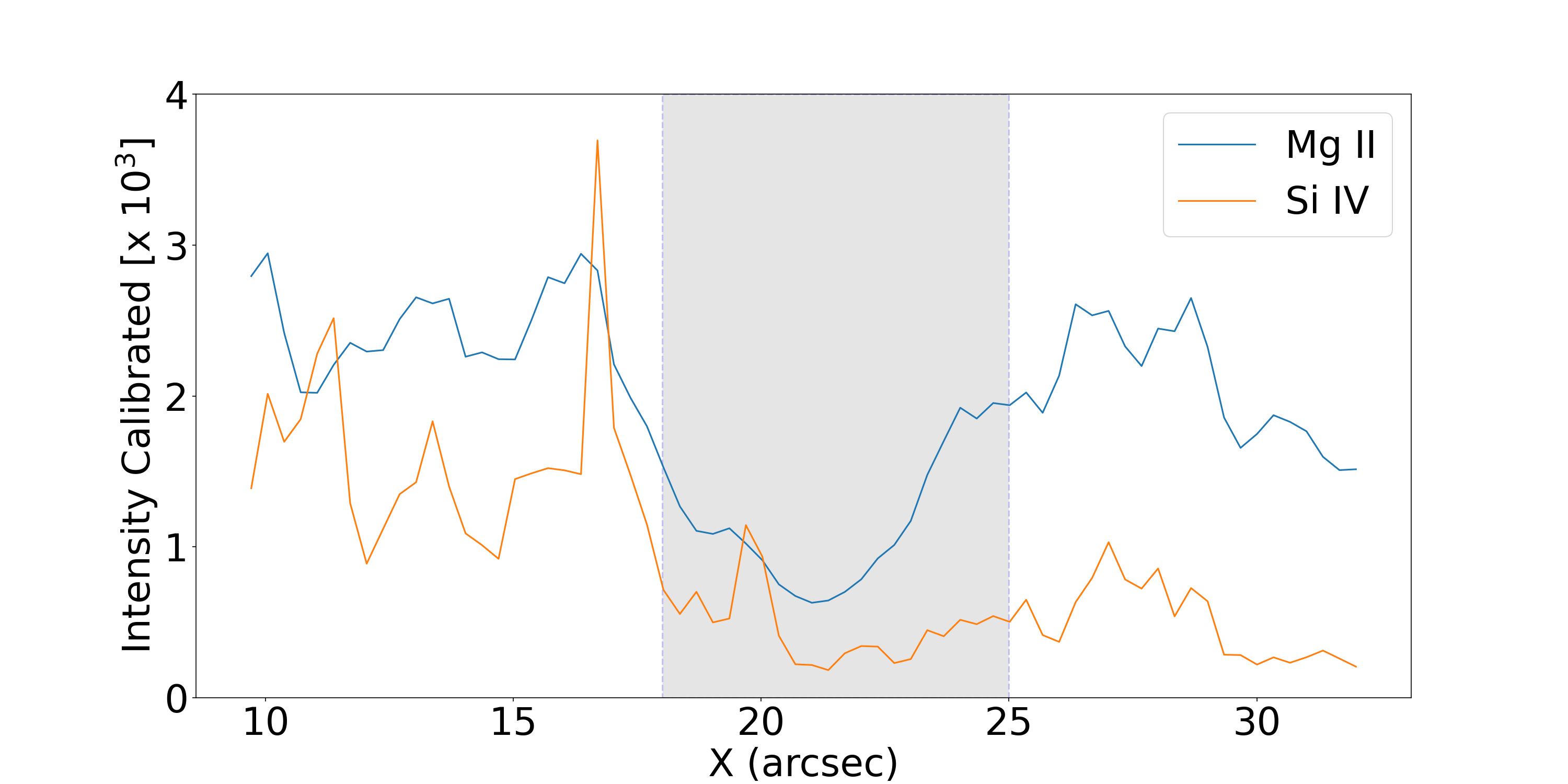}
\includegraphics[scale=0.2,trim=30 30 60 20]{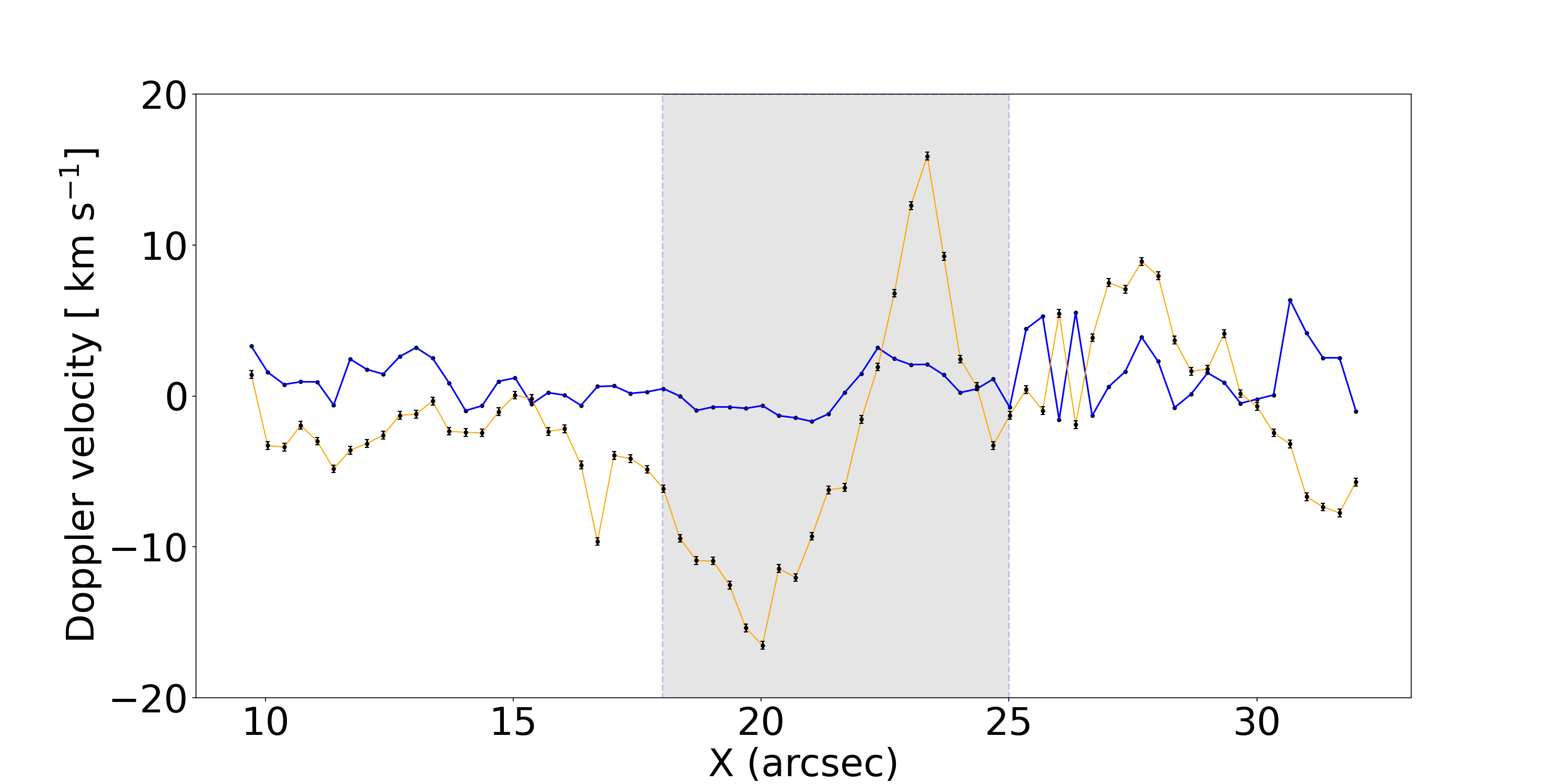}
\caption{ \mg and \si intensity and velocity variation along the slit drawn in the third middle panel of Fig. \ref{Fig7b}. The shaded area indicates the width of the filament as seen in the \mg intensity maps. Error bars in the \si velocity plot are retrieved as explained in Sect. \ref{UVproc}}
\label{Fig12}
\end{figure*}

\begin{figure}
\centering
\includegraphics[scale=0.5, clip, trim=20 417 0 200]{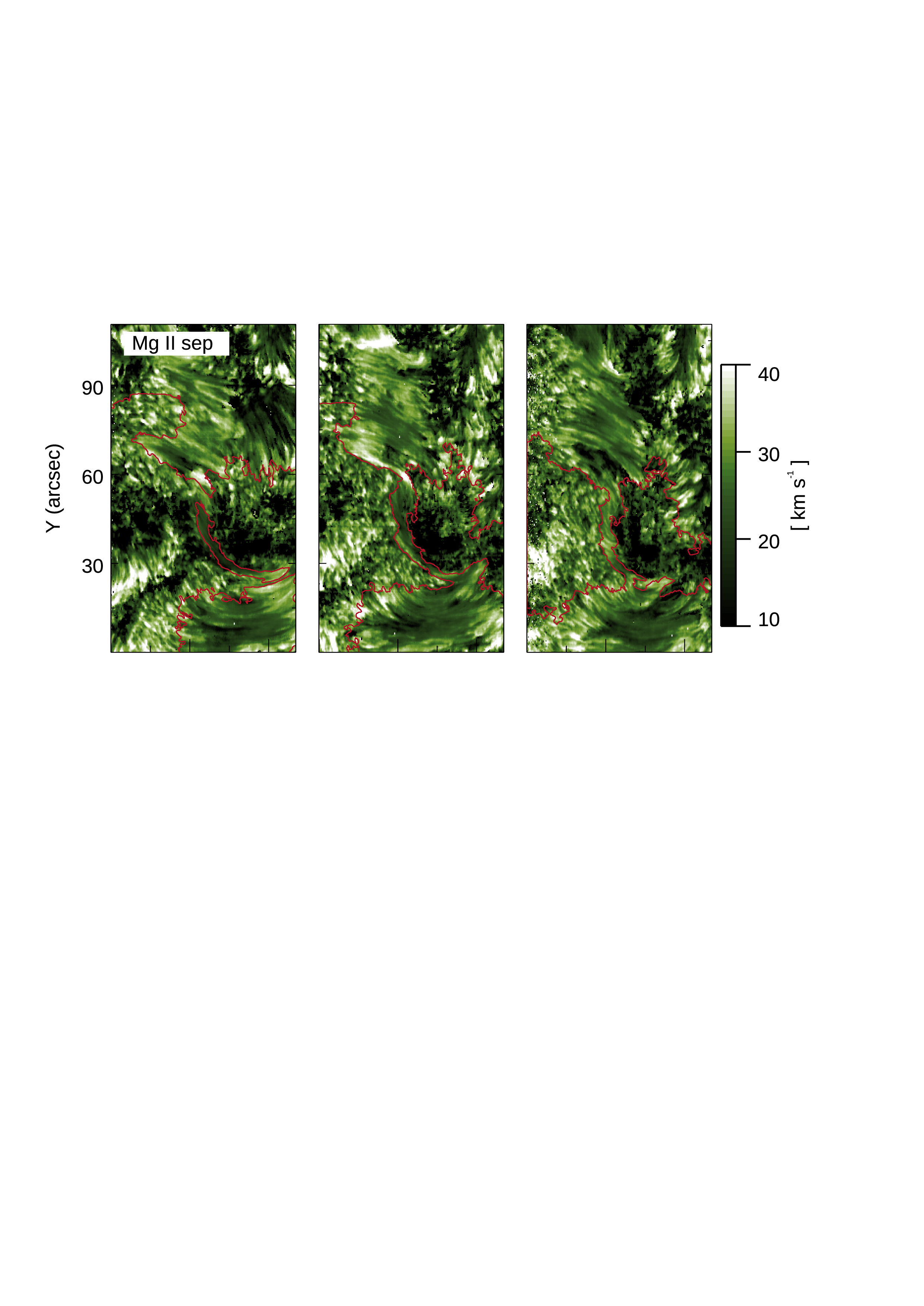}\\
\includegraphics[scale=0.5, clip, trim=20 417 0 200]{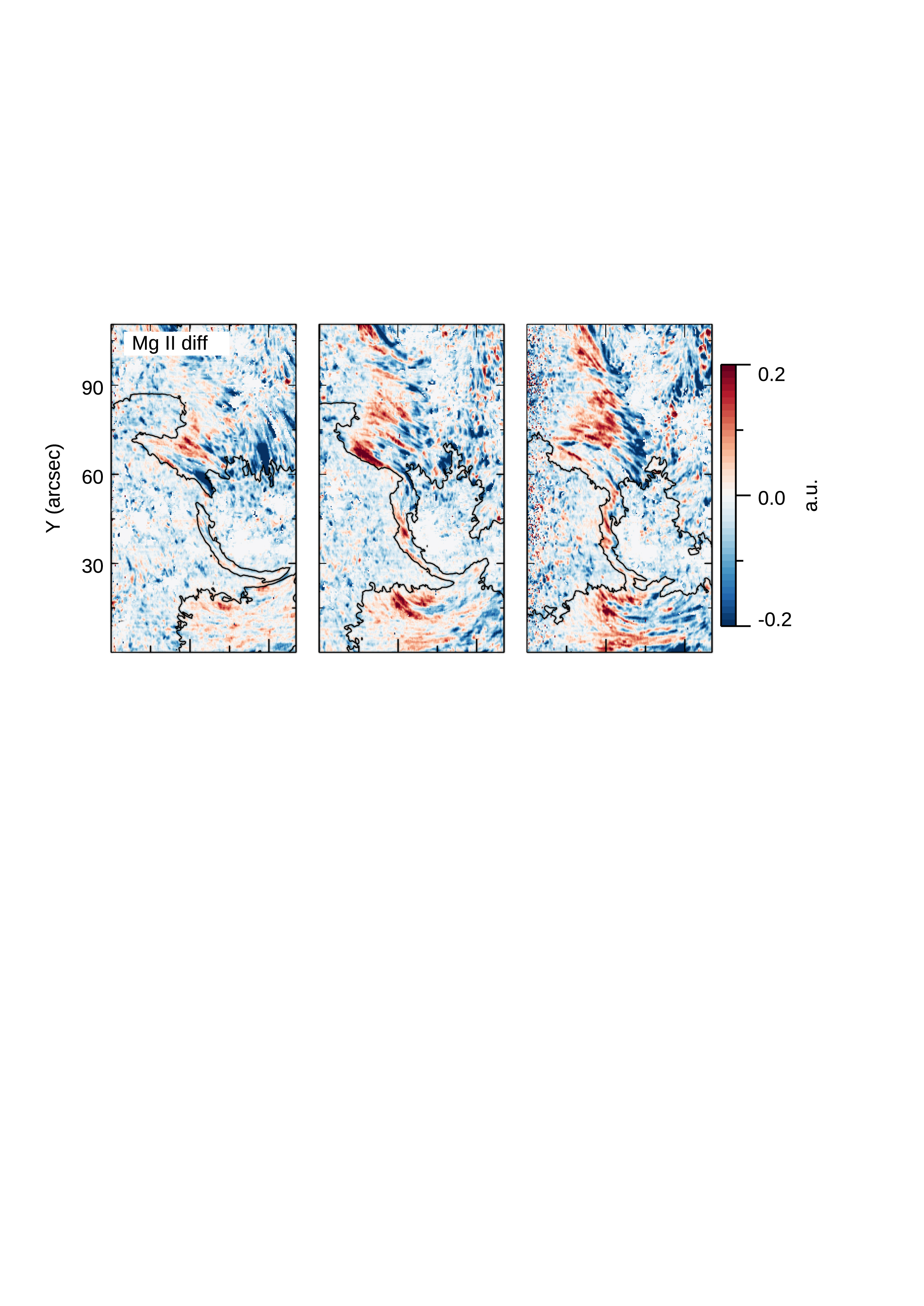}\\
\includegraphics[scale=0.5, clip, trim=20 417 0 200]{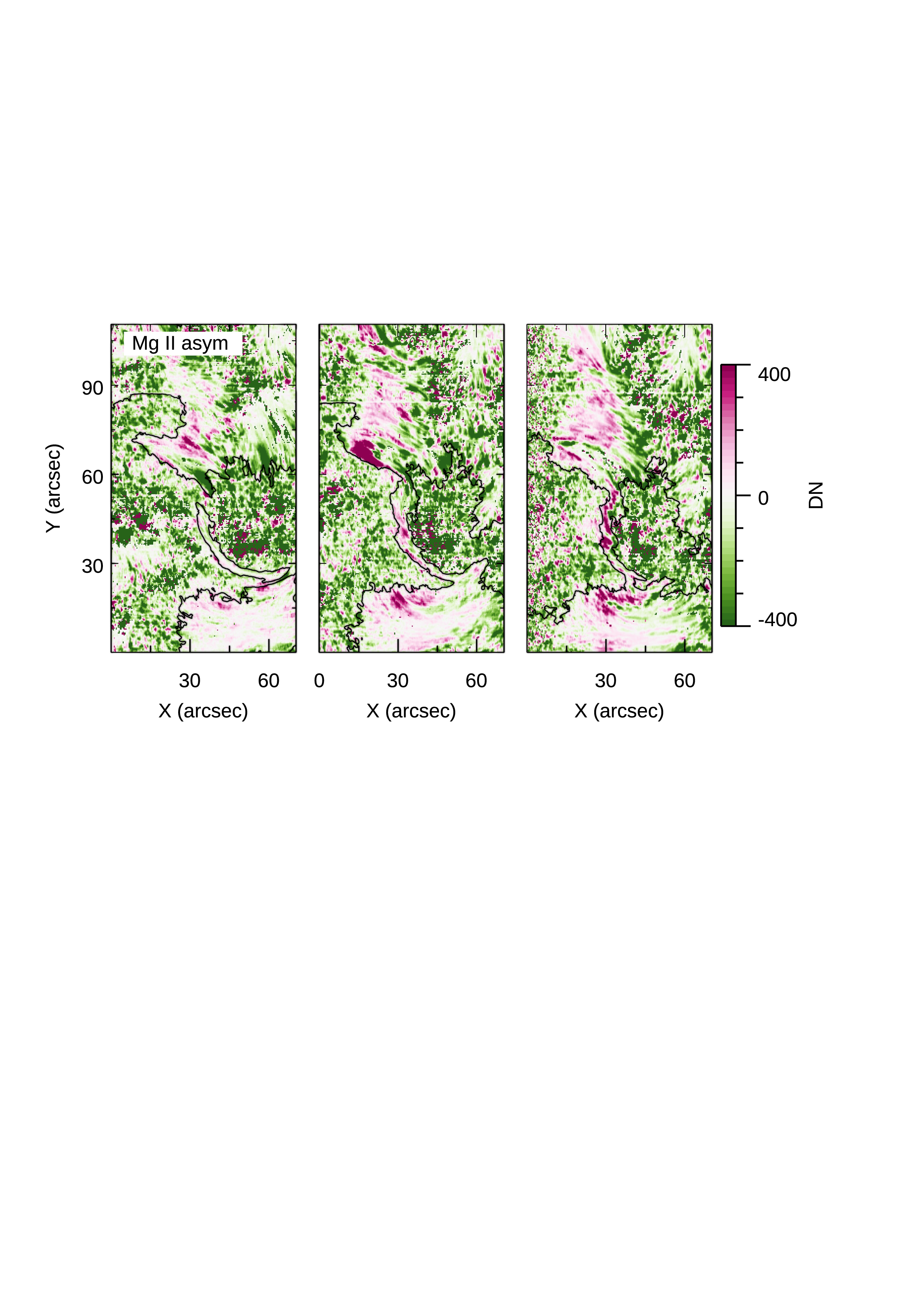}\\
\includegraphics[scale=0.5, clip, trim=20 375 0 200]{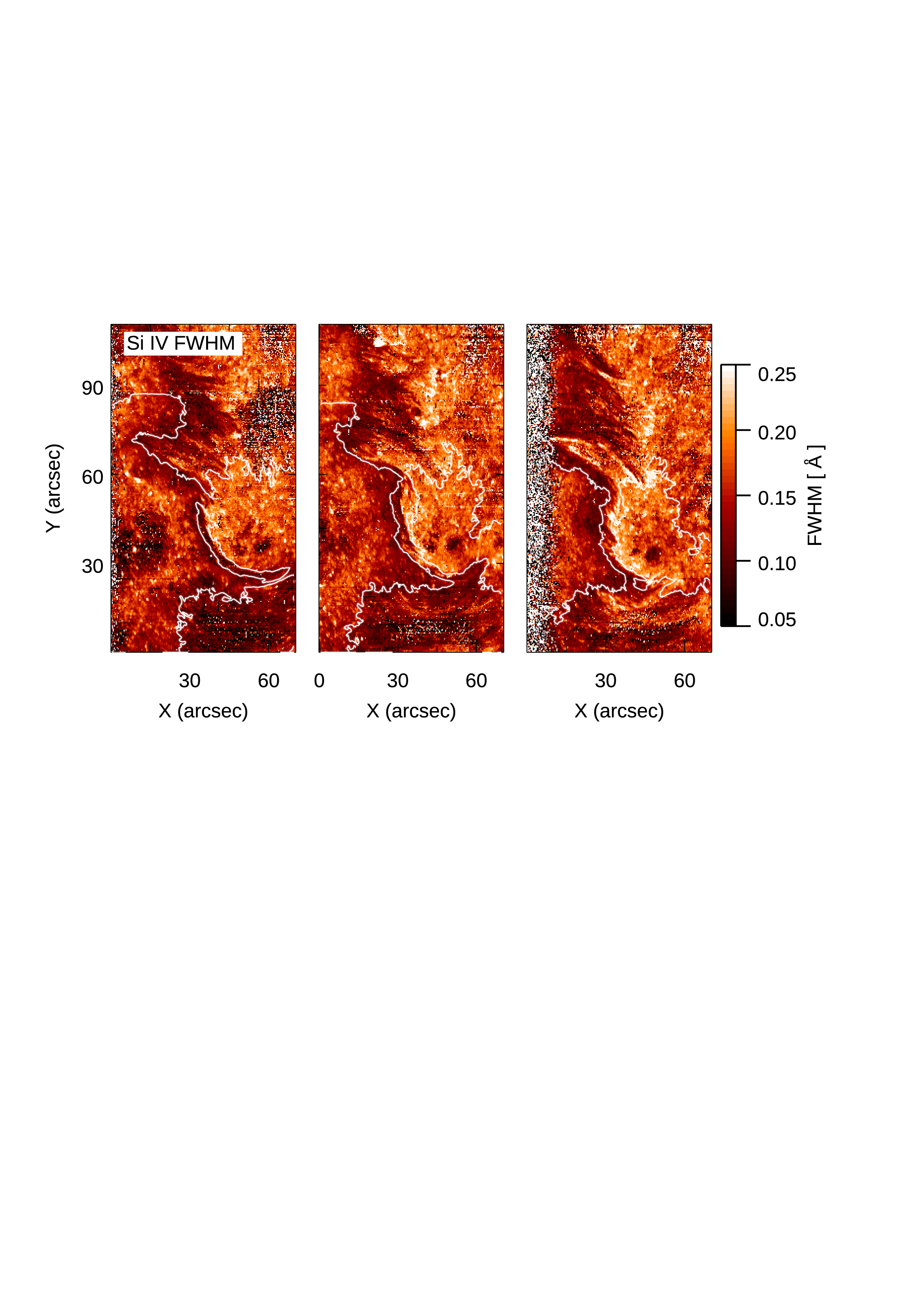}\\
\caption{From top to bottom: Mg II peak difference (diff), peak asymmetry (asym), peak separation (sep) and \si FWHM maps. Red and black contours indicate Mg II k peak intensity in all panels.}
\label{Fig7}
\end{figure}

\subsection{UV Doppler velocities and velocity gradients}

Doppler velocity maps for the three IRIS lines are displayed in Fig. \ref{Fig6}. As in Fig. \ref{Fig5}, the maps are organized from top to bottom by spectral line: \mg k$_{3}$, \ion{C}{ii} and \si, respectively, for the three IRIS raster scans. In all panels, a smooth contour outlines the filament. This contour was defined by applying an intensity threshold to the \mgk peak intensity line and then it was cleaned up to highlight the filament structure as a whole.

In all maps, the filament is blue-shifted. As has long been observed and described \citep{Parenti2014,Sergio18}, a clear flow of material is present along the AFSs, moving from one footpoint to the other through the system of bundles. This feature becomes particularly evident during the last two scans (11:03–11:52 UT and 20:48–21:37 UT, shown in the middle and right panels of Fig. \ref{Fig6}). The plasma patterns observed in the AFSs correspond closely to the representation given in Figure 15 of \citet{Sergio18} and have been observationally reported in \citet{Sergio20} and more recently in \citet{Reetika2024}. Notably, in the middle raster scan, only the \mg k$_{3}$ Doppler velocity map clearly reveals the presence of the redshifted area at the tail of the two filament sectors previously identified in H$\alpha$ and \ion{He}{i} images, with velocity exceeding 15 km s$^{-1}$. Furthermore, this red-shifted footpoints are barely visible in the \si map. This may suggest that the northern AFS has already risen to higher layers of the atmosphere. A similar behavior is shown in panel (e) of Figure 15 in \citet{Sergio18}, where \ion{He}{i} observations of their AFS exhibit gradually decreasing downflows at the footpoints. The three Doppler velocity maps display a different range of values. In the \mg k$_{3}$ Doppler velocity maps, the filament shows upward velocities in the range of [-0.5,-3] km s$^{-1}$.
The \si maps report relatively higher velocities in the filament, with values in the range of [-6,-15] km s$^{-1}$. 
The \ion{C}{ii} Doppler velocity maps are noisier compared to the other two UV maps. As a result, the velocity pattern within the AFS is less clear compared to the \mg k$_{3}$ and \si\ maps. The maps clearly show that inside the \mg k$_{3}$ contour, the plasma consistently exhibits blueshifts across all three datasets, indicating coherent upward motions in this region despite the increased noise in the \ion{C}{ii} observations.

By closely inspecting the IRIS \mg k and \si peak intensity and velocity maps, we can compare the fine structure of the filament in these spectral lines and examine the relationship between intensity and Doppler velocity. Figure \ref{Fig7b} displays three different zoomed images, outlined by red boxes in the \mg k peak intensity map of Fig. \ref{Fig5}, as seen in the \mg k and \si1393 \AA\ peak intensity, along with the corresponding H$\alpha$ sections for comparison. Note that the H$\alpha$ image corresponds to the last available frame taken at 10:58 UT before the IRIS observation, while the IRIS rasters were acquired between 11:03 and 11:52 UT. Although the filament is clearly visible in the core intensity maps of both \mgk\ and \si, its fine structure is better resolved in the H$\alpha$ observations. Nevertheless, the figure provides useful insight into the spatial relationship between the filament intensity and the surrounding Doppler velocity field. Indeed, Figure \ref{Fig12} shows the intensity and Doppler velocity profiles along the black horizontal segment indicated in all panels of the middle row in Fig. \ref{Fig7b}. In this horizontal cut, a distinct drop in intensity marks the position of the filament. Blueshifted plasma is associated with this intensity dip, as particularly evident in the \si data. Although the intensity decreases on both the eastern and western sides of the filament, it does not recover to pre-filament levels on the western side, suggesting additional structuring or obscuration. Moreover, positive Doppler velocities are observed just before the western edge of the filament, indicating localized downflows in that region. This effect may be related to the filament's position on the solar disk. In addition, compared to previous figures, the \si maps highlight more clearly that the redshifted footpoint visible in the \mg k$_{3}$ velocity does not appear as prominently in the \si Doppler data, further supporting the idea that the downflow is confined to lower atmospheric layers.

Figure \ref{Fig7} reports \mg k$_{2}$ peak separation, difference and asymmetry maps. These maps display patches of bad calculation on the eastern and western area at the edges of the filament due to single peaked profiles (k$_{2R}$ or k$_{2B}$ not present, as already shown in the top-right panel plot of Fig. \ref{cal1}). 
The filament displays a k$_{2}$ peak separation that increases with time (see the top panels of Fig. \ref{Fig7}). Indeed, in the third IRIS scan, we observe inside the filament a k$_{2}$ peak separation larger than 30 km s$^{-1}$. Concerning the k$_{2}$ peak difference, shown in the second-row panels of Fig. \ref{Fig7}, the filament is mainly characterized by a negative ratio, indicating stronger blue peaks, corresponding to upflowing material above the \mgk peak formation height. A positive ratio (with red peaks stronger than blue ones) appears to move along the length of the filament during the two last IRIS scans indicating downflowing material. Moreover, these two last scans display a higher concentration of \mg profiles with positive k$_{2}$ asymmetry (i.e., upflows) inside the filament k$_{3}$ contour. Some patches of negative k$_{2}$ asymmetry (i.e., downflows) are also visible.

Fig. \ref{Fig7} bottom panels display \si FWHM maps. 
On the western side of the FoV we observe broader \si profiles and this characteristic increases with time, reaching a FWHM value of up to 0.25 \AA\ in the third IRIS scan. 
As can be noted, this pattern seems to be persistent in time. The profiles appear to be broadened twice as much with respect to the eastern side.  It can easily be possible to recognize three regions having different values of FWHM. In particular, the eastern side of the filament is characterized by a FWHM $<$0.15 \AA, while we find values of FWHM between 0.1 and 0.2 \AA\ inside the filament, and the values are greater than 0.2 \AA\ on the western side of the filament. Knowing the IRIS instrumental and thermal broadening (see \citealp{DelZanna2018}), the FWHM for an optically thin line (as the \si) is related to the non-thermal velocity. Assuming a temperature formation of 80 $\times$10$^{3}$ K for the \si 1393.755 \AA\ line, a thermal width of 6.88 km s$^{-1}$ and the instrumental broadening equal to 3.9 km s$^{-1}$ \citep{depontieu2014}, then, the filament has a non-thermal velocity between 13 km s$^{-1}$ and 29 km s$^{-1}$, while on the eastern and western sides of it we find values lower than 21 km s$^{-1}$ and greater than 29 km s$^{-1}$, respectively. The values reported for QS and AR areas using SUMER \si data are on the order of 27 km s$^{-1}$, which are comparable to those found on the western side of the filament. For temperatures with logT$<$5.4, as in the case of \si data, \citet{Parenti2007} reported lower non thermal velocity values in prominences compared to those in the QS. 

\section{Discussion}

\begin{figure*}
\centering
\includegraphics[scale=0.45,trim=0 80 0 50, clip]{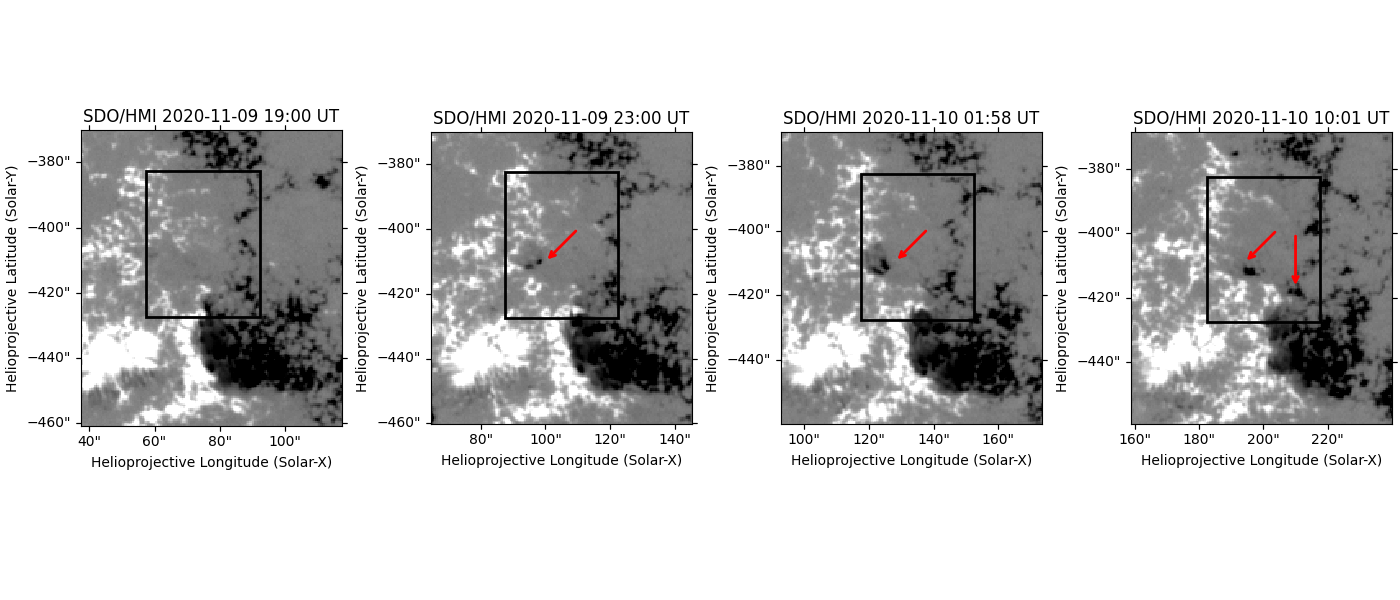}
\includegraphics[scale=0.45,trim=0 80 0 50,clip]{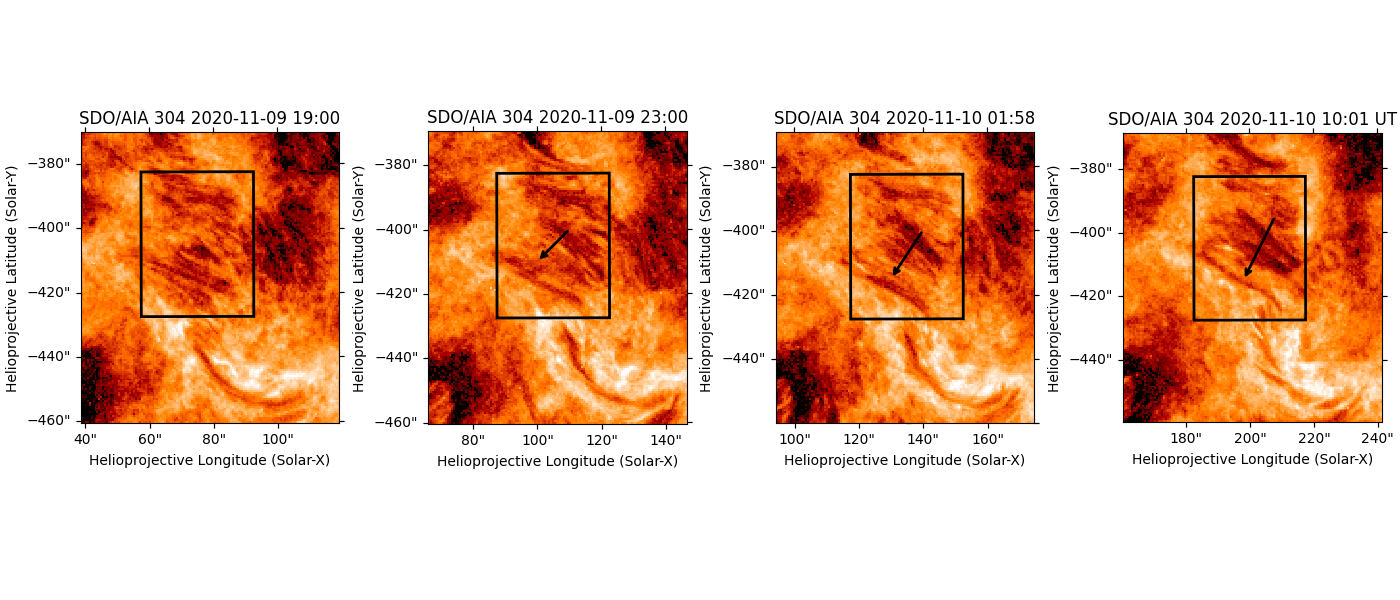}
\caption{SDO/HMI magnetograms (top panels) and SDO/AIA 304 \AA\ images at four representative times. The black boxes indicate the magnetic configuration of the AFS. Magnetograms are clipped $\pm$ 500 G. Red and black arrows point new magnetic features described in the text.}
\label{Figc}
\end{figure*}
 
In this paper, we presented a multi-wavelength, multi-instrument observation of an AR filament. These observations are especially noteworthy due to the combination of high-spatial resolution H$\alpha$ images and high-spectral resolution \ion{He}{i} data from the GREGOR telescope, complemented by near-simultaneous UV IRIS spectra and images sampling the solar chromosphere and transition region. This comprehensive data set enabled us to investigate the filament’s fine structure and plasma dynamics with unprecedented detail.
 
The high-spatial resolution H$\alpha$ observations reported in this work reveal that the filament thickened over time (10:54 UT on November 9 and 10:48 UT on November 10), becoming darker and more compact. The fibril patterns became more pronounced (the online movies and Fig. \ref{fig4b}).

We observed a change in the orientation of the threads over a time span of almost twelve hours, which is consistent with the timescale reported in the literature. 
Although we observe H$\alpha$ fibrils changing orientation, we do not capture in this spectral range any merging episodes or small-scale activity indicating interactions of threads and subsequent thickening of the filament.  
On November 10, H$\alpha$ observations show that the filament appears darker and more compact compared to the previous day. A new, darker, and more compact structure also appears in the northern part of the FoV, resembling an extension of the filament. The simultaneous GRIS \ion{He}{i} raster scan never clearly shows the presence of the main filament along its full length (Fig. \ref{fig4b}) possibliy reflecting variations in the PCTR across different segments of the filament, differences in the local EUV illumination or a combination of both effects. The helium core exhibits several thin threads in the region where H$\alpha$ displays darker structures. 
Stokes-I profiles in the central part of the filament are shallower and reduced of over 30\%. Furthermore, this central part has optical thickness values below unity compared to the darker part of the filament and the thin threads, whose values exceed 1.5 (Fig. \ref{fig4bis}). Small-scale brightening events are observed in the \ion{C}{ii} 1335 \AA\ and \si 1393 \AA\ peak intensity maps. The presence of threads in between the observed brightening events, may suggest possible thread formation/disappearance, as reported in \citet{Zhou2016}.

As mentioned in Sect. \ref{sec1}, most studies of plasma flows in the UV range have focused on prominences or filaments observed near the limb. In contrast, our analysis concerns a filament located well on the solar disk, where projection effects are minimized ($\mu$ = 0.78–0.66). We report that the filament is consistently blue-shifted in all the analyzed spectral lines (\ion{He}{i}, \mgk, \ion{C}{ii}, and \si) throughout the entire observation period. These blueshifts, interpreted in terms of upward velocities, highlight the presence of persistent upflows even in on-disk observations, as also reported in previous works \citep{Schmieder1984, Kucera1999}. Extending the study of \citet{Schmieder1984}, where the authors reported blue-shifts using {H$\alpha$ and the transition region \ion{C}{iv} 1548 \AA\ spectral lines on the order of [0.5, 5.6] km s$^{-1}$}, here are able to infer Doppler velocities by using the IR \ion{He}{i} 10830 \AA\ and the three UV \mgk, \ion{C}{ii} 1335 \AA\ and \si 1393 \AA\ lines, all of them having persistent blue-shift velocities on the order of [0.5, 15] km s$^{-1}$. Noticeably, the IR \ion{He}{i} and UV \mg k$_{3}$ Doppler velocity maps highlight red-shifted signatures at the end of the new dark and compact structure visible during the co-temporal ground-based and IRIS observations. This redshifted feature is barely visible in the \si map. The observed redshifted structure may be associated with the footpoints of the northern AFS and possibly also with those of the filament, reflecting an overlap between two structures in the chromosphere. This association can be seen in Fig. \ref{Figc}, which shows zoomed-in SDO/HMI and SDO/AIA 304 \AA\ images. Indeed, the northern AFS displays a clear magnetic configuration in the SDO/HMI data, highlighting its connection between opposite polarities, while the filament is located along the PIL. 

Furthermore, in the images from 23:00 UT on November 9 (fig. \ref{Figc}, see arrows), the new, compact, and darker structure, also visible in H$\alpha$ and \ion{He}{i} on November 10 (fig. \ref{Fig7b}, top panels), displays the magnetic configuration of a filament, i.e. a feature along the PIL. This supports the probable coexistence of the northern AFS and of this new filament in the \ion{He}{i} and \mgk$_{3}$ velocity maps. The reduced magnitude of the \si Doppler velocity may indicate that redshift signal relative to the northern AFS  does not reach higher atmospheric layers. We recall that the estimated errors for the \si velocities are comparable with previous studies, being of the order of 1 km s$^{-1}$. 

The study performed by \citet{Kucera1999} showed plasma upflows in the H$\alpha$ observations acquired with the ground-based VTT telescope, but they could not detect any blue-shift in the SUMER \si transition region line. The Doppler velocity retrieved using the IR \ion{He}{i} spectral line displays a clear blue-shift velocity in the filament in the range of [0.5,3.5] km s$^{-1}$ with an average Doppler velocity of $-1.5$ km s$^{-1}$. The spectral range of \ion{He}{i} 10830 \AA\ was used by \citet{Kuckein2012b}, who reported the study of Doppler velocities in an AR filament. In particular, they found that the filament's spine is dominated by upflow (reaching values of about -1 km s$^{-1}$). On the contrary, the section of their filament observed above the orphan penumbra in the chromosphere appears to be redshifted, suggesting a different mechanism to sustain or create the filament (from below the photosphere in their case). \citet{Baso19} studied an AR filament located in a flaring area on chromospheric GRIS \ion{He}{i} data, and reported upflow along the PIL of the order of -3 km s$^{-1}$. Unlike previous studies that reported a mix of upflows and downflows in different parts of AR filaments, our observations reveal consistently blue-shifted velocities along the filament, across all time windows and spectral ranges. While such differences may be due to variations in magnetic configuration, evolutionary stage, or observational setup, the coherence observed here could reflect the advantage of combining high-cadence, multi-wavelength data over extended periods.

In addition, we report the \mg k$_{2}$ separation, k$_{2}$ difference and k$_{2}$ asymmetry. These quantities are usually investigated in flare studies, such as in \citet{Polito2020,Polito2023a,Polito2023b}. Together with the \mg k$_{3}$ Doppler shift, these quantities are good diagnostics of plasma flow and velocity gradients in the solar chromosphere \citep{leenaarts2013}.  
In our study, their values point to the presence of plasma flow and gradients along the full filament length. It is worth noting that the correlations shown in \citet{leenaarts2013} and \citet{Pereira2013} have some spread, and they are valid for spectra at disk center and cover QS conditions to which the simulation belongs. Since the IRIS launch, the \mghk lines have also been modeled for different magnetic field topologies, such as those representative of QS, ephemeral flux regions and plages \citep{Hansteen2023}, as well as under various conditions like flares \citep[e.g.,][]{Kerr2016} and in prominence \citep{Heinzel2014,Heinzel2015}. Although these studies are not properly representative of our observations, the asymmetrical profiles of \mg seem to be a result of fine unresolved plasma motion \citep{Peat2023}. Further dedicated numerical studies are needed to better understand the \mg asymmetric profiles and the associated plasma dynamics in active region filaments, as well as their variation across the solar disk.

At both the eastern and western sides of the filament, but more clearly seen on the western side, a transition in plasma flow is evident, characterized by blue-shifts within the filament and red-shifts outside of it. This kind of behavior has previously been pointed out by \citet{Engvold1985}. 
AR filaments showing chromospheric upflows (using the IR \ion{He}{i}  spectral line) are embedded in a downflow area corresponding to the faculae region \citep{Kuckein2012b,Xu2012} and moss region \citep{Zhao2024}. \citet{Kuckein2012b} speculated that such a downflow at the faculae region could probably be associated with coronal rains and/or their different manifestations \citep{Lagg2007}. The results presented in \citet{Zhao2024} display brightening events and apparent downflow in the moss region (footprints of 2-3 MK loops beneath the 1 MK loops). 

Finally, the non-thermal velocities reported in our case are comparable with those found in QS and AR by \citet{Teriaca1999}, whole being lower than those in the QS inside the filament as reported in \citet{Parenti2007} in the temperature range analyzed (logT$<$5.4).

\section{Conclusion}

In summary, our multi-wavelength analysis provides a coherent view of the plasma dynamics in an AR filament, from the chromosphere to the transition region. By combining IR and UV Doppler diagnostics, we tracked persistent flow patterns over two consecutive days, revealing a stable velocity field across a wide range of temperatures. 

Indeed, throughout the observing period, we consistently detected blue-shifted velocities in the \ion{He}{i}, \mg k$_{3}$, \si and \ion{C}{ii} lines, indicating stable upflows from the chromosphere into the transition region. In contrast, red-shifted features at the footpoints of a newly formed structure, clearly visible in the \ion{He}{i} and \mg lines, suggest chromospheric downflows possibly due to an overlap with an AFS. The absence of this signal in \si may point to its confinement to lower atmospheric layers. Additionally, asymmetries in the \mgk$_{2}$ support the presence of flows and velocity gradients along the entire filament spine.

Importantly, this work represents one of the very few studies to combine ground-based IR observations of \ion{He}{i} 10830 \AA\ with UV transition region diagnostics (e.g., \mg, \si and \ion{C}{ii}) for an AR filament. While previous studies have primarily focused on quiescent filaments or prominences, often relying on indirect velocity measurements, our analysis directly captures dynamic behavior across multiple atmospheric layers with substantially improved resolution. In doing so, we confirm and extend early findings from missions such as SMM and SUMER, while addressing an area of solar physics that has remained relatively unexplored in the high-resolution era.

Despite known limitations in the IRIS dataset, our results present a consistent and robust picture of the filament dynamics across multiple layers of the solar atmosphere. This demonstrates the potential of coordinated IR and UV spectroscopy for studying fine scale filament evolution and plasma motions. Future observational proposals could take advantage of the lessons learned from this study by integrating both high-resolution chromospheric observations from instruments such as CRISP \citep{CRISP}, CHROMIS \citep{CHROMIS}, and GRIS and those available at the DKIST telescope, and coronal observations from Hinode/EIS \citep{EIS}, and SO/SPICE \citep{SPICE}. More systematic and continuous observational campaigns, targeting a filament over several days and utilizing complementary instruments capable of resolving both the plasma flows and magnetic field structures, would allow us to infer all relevant physical parameters such as mass density, temperature, velocity and magnetic field. This approach could pave the way for future synergistic studies combining spectropolarimetric measurements from IBIS2.0 \citep{IBIS2.0:20,IBIS2.0:22,IBIS2.0:24} 
in the photosphere and chromosphere with UV observations from Solar-C \citep{solarc19} and MUSE \citep{muse2020,muse2022}, offering a comprehensive understanding of the filament's kinematic and magnetic properties.

\begin{acknowledgements}
This research was supported from the European Union's Horizon 2020 research and innovation programme under grant agreement no.~824135 (SOLARNET) and by the Italian Space Agency (ASI) under contract with the INAF no.~2021-12-HH.0 and Addendum 2021-12-HH.1-2024 “Missione Solar-C EUVST–Supporto scientifico di Fase B/C/D”, and under contract with the INAF no.~2022-29-HH.0 “Missione MUSE”. MM has been also supported by the ASI-INAF agreement n. 2022-14-HH.0 and I/028/12/7-2022. 
CK acknowledges grant RYC2022-037660-I and SJGM
grant RYC2022-037565-I, both funded by MCIN/AEI/10.13039/501100011033 and by "ESF Investing in your future". Financial support from grant PID2024-156538NB-I00, funded by MCIN/AEI/ 10.13039/501100011033 and by “ERDF A way of making Europe” is gratefully acknowledged by SJGM."
MM would like to thank Ilaria Ermolli and Roberto Piazzesi for their valuable discussions, insightful suggestions, and assistance with language editing, all of which contributed to improving the manuscript.
\end{acknowledgements}

%
%

\clearpage
\begin{appendix}
\newpage
\section{\mg processing }\label{mgd}
The tool used to retrieve the characteristic of the \mg spectral line presents some limitations because it works well for double-peaked or strongly shifted single-peak profiles (i.e., not for the optically thin regime). Furthermore, the k$_{3}$/h$_{3}$ properties are set to NaN when the result is believed to be unreliable. This is also the case for the peak properties, the result is a more ‘dark noise’ in the maps (see for more detail the ITN26\footnote{\url{https://iris.lmsal.com/itn26/iris_level2.html}}.)

The diff, the sep and asym quantities are derived from the k$_{2}$ and k$_{3}$ values using the following formulae \citep[e.g.,][]{Polito2020,Polito2023a,Polito2023b}: 

\begin{equation}
	\mathrm{diff} = \frac{I_{k_2,R} - I_{k_2,B}}{I_{k_2,R} + I_{k_2,B}}
\end{equation}

\begin{equation}
	\mathrm{asym} = I_{k_2,R} - I_{k_2,B}
\end{equation}

\begin{equation}
	\mathrm{sep} = v_{k_2,R} - v_{k_2,B}
\end{equation}

\noindent
where I/v$_{k_2,B}$ and I/v$_{k_2,R}$ are the intensity and Doppler velocities of the blue and red peaks of the \mgk lines, respectively.
 
\section{\si and \ion{C}{ii} absolute calibration and errors}
\label{si}

\subsection{Absolute calibration procedure}
\begin{figure}
	\centering
	\includegraphics[scale=0.17,trim=90 0 60 
	0]{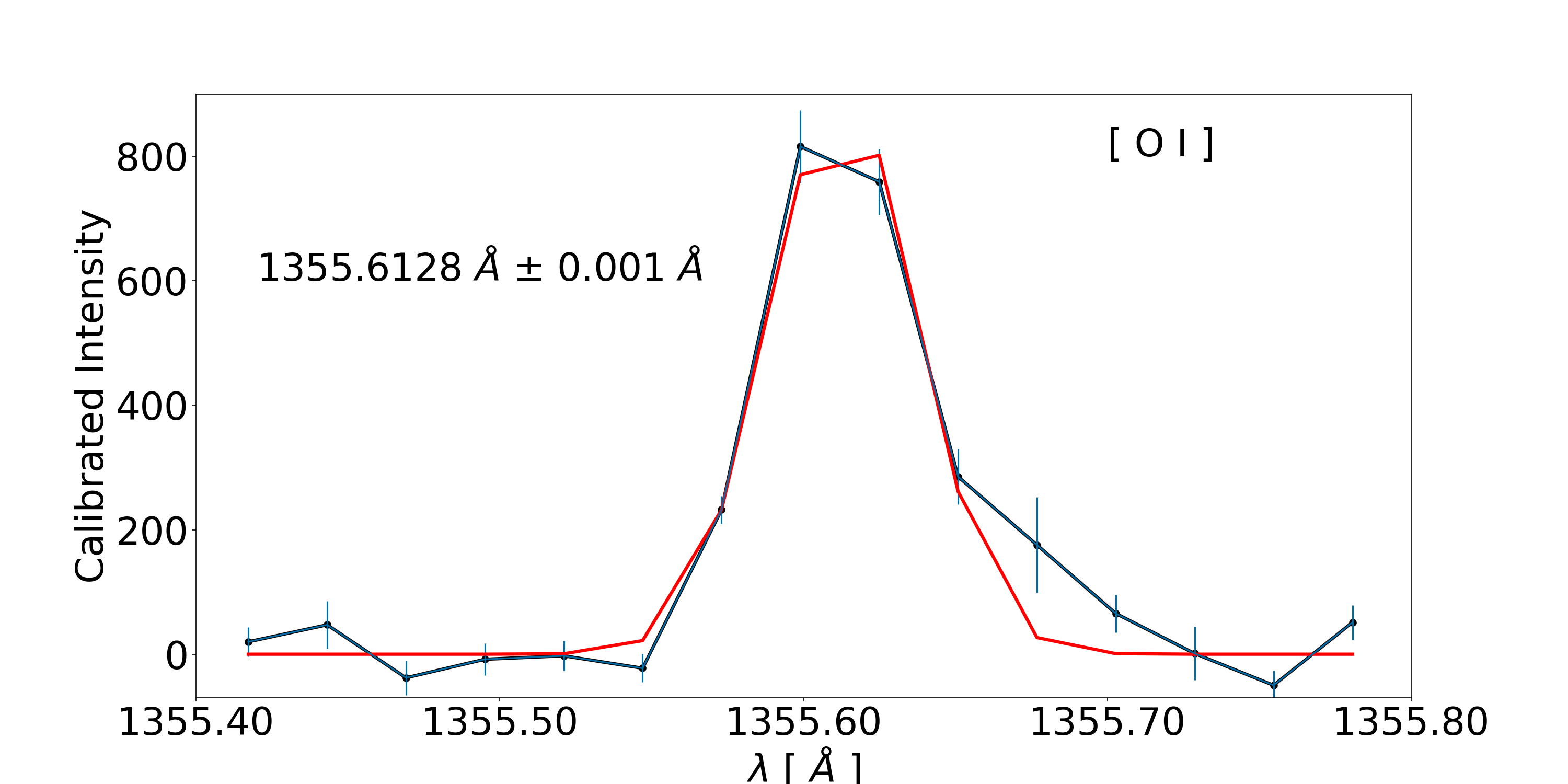}
	\includegraphics[scale=0.17,trim=90 0 60 
	0]{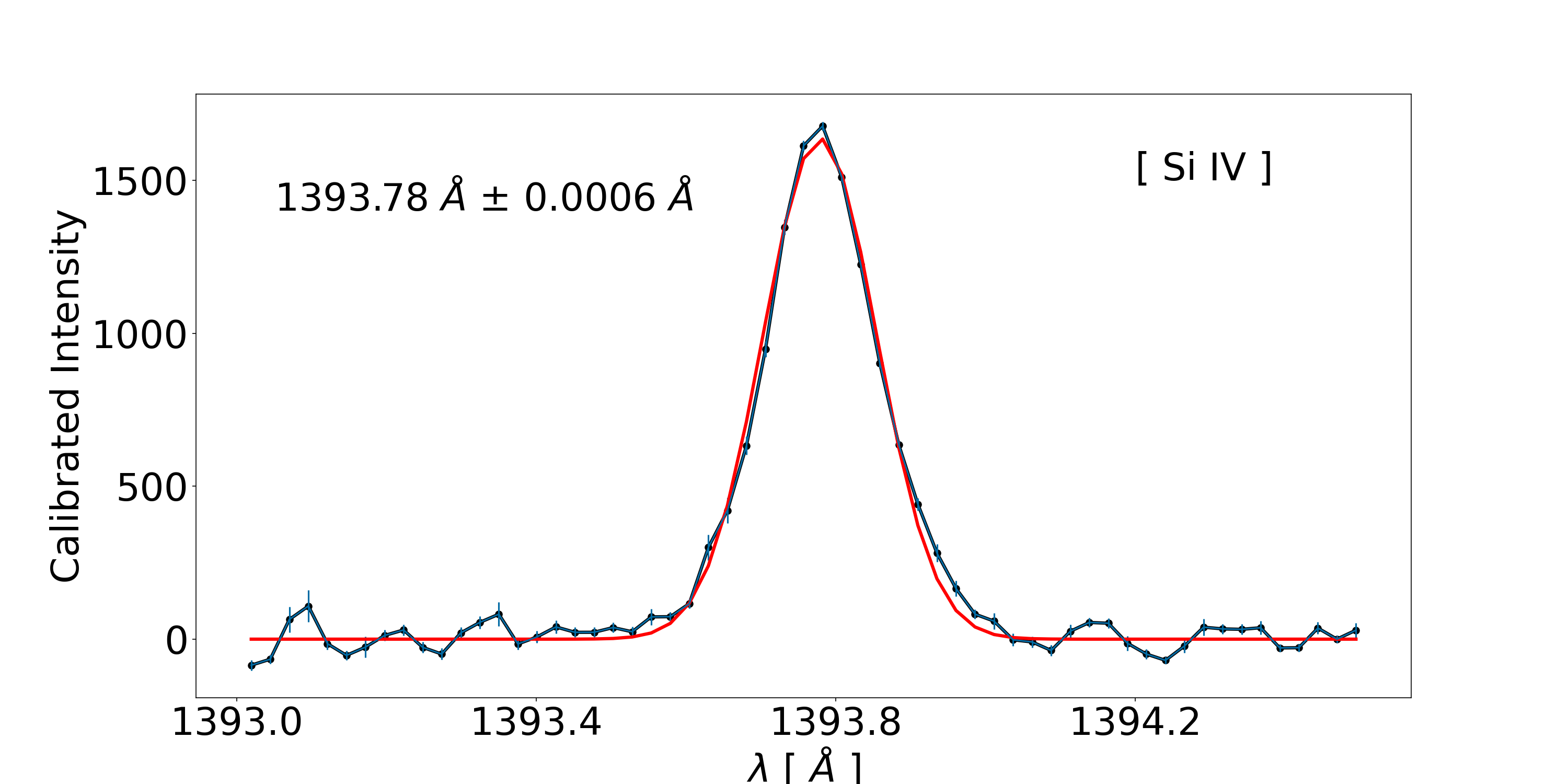}
	\includegraphics[scale=0.17,trim=90 0 60 
	0]{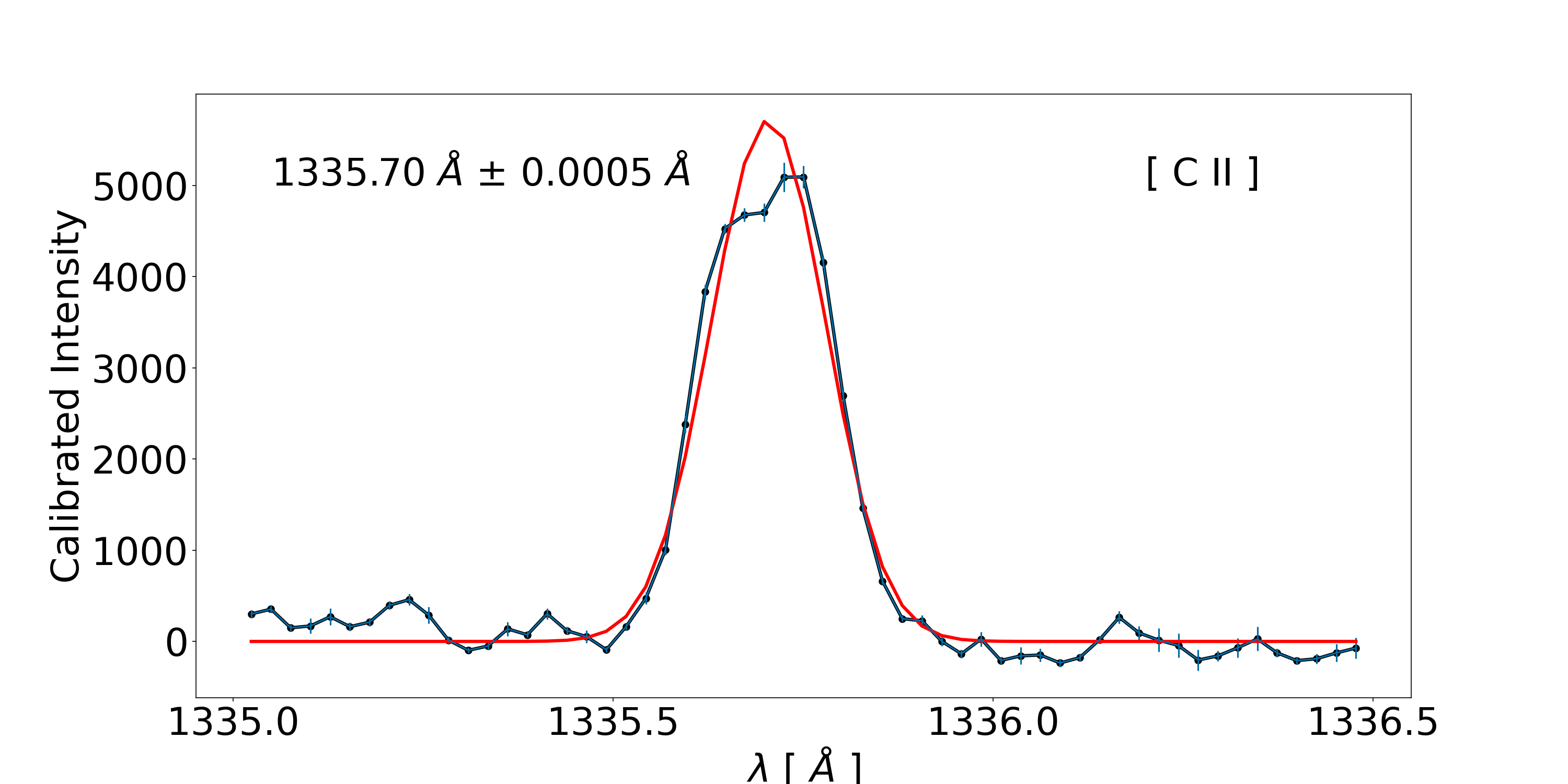}
	\caption{From top to bottom: \ion{O}{i} 1355.6 \AA , \si 1393 \AA\ and \ion{C}{ii} 1336 \AA\ average spectral profiles. The observed spectral lines (black curves) are averaged over the entire QS FoV taken on 12 November 2020. The best fit with a Gaussian is shown in red. Error bars on the black curves represent the standard deviation of the intensity at each wavelength point, computed across all spatial pixels in the QS FoV. Intensities are given in units of [$erg$ $cm^{-2}$ $s^{-1}$ $sr^{-1}$ $Angstrom^{-1}$].} 
	\label{cal}
\end{figure}

The ITN20\footnote{\url{https://iris.lmsal.com/itn20/iris_level2.html}} document suggests to follow different strategies to perform absolute calibration depending on the FUVs and NUV spectral ranges. In the case of FUVL wavelength range (1389 - 1407 \AA) the \ion{Fe}{ii} 1392.8 \AA\ spectral line it is usually chosen as a reference of zero shift. In our observing mode this spectral line is very weak and thus we decide to use the \ion{O}{i} 1355.6 \AA\ in the FUVS range instead. The FUVL detector has a fixed wavelength offset from the FUVS one. This is within a small fraction of pixels or $\pm$0.5 km s$^{-1}$ (defined as $\sigma_{f}$) \citep{Wulser2018}. Then, we use the \ion{O}{i} 1355.6 \AA\ line for both ranges to perform an absolute calibration. For this purpose, we use a QS observation taken at the disk center on November 12 with the same observing mode (dense 320-step raster) and the same exposure time of our observations. Figure \ref{cal} displays the average spectra of \ion{O}{i} 1355.6 \AA\, \ion{Si}{iv} and \ion{C}{ii} lines (black) obtained over the entire IRIS QS raster FoV. The black solid lines are the original line profile, whereas the overplotted red lines the corresponding Gaussian fit. The wavelength at rest measured in the laboratory for the \ion{O}{i} line is assumed to be 1355.5977 \AA\ \citep{Eriksson68}.

\subsection{Error Calculation Procedure}

The velocity measurements arise from the retrieved Gaussian fit parameters. In particular the parameter A$_{1}$ representing the Gaussian center was used to derive the velocity. 
Thus, the total uncertainty for the velocity is estimated as follow:

$\sigma_{v}^{2}$=$\sigma_{r}^{2}$+$\sigma_{c}^{2}$+$\sigma_{xy}^{2}$+$\sigma_{f}^{2}$

\noindent
where, the term $\sigma_{r}$ accounts for the uncertainty in the reference wavelength and is equal to 0.13 km s$^{-1}$. $\sigma_{c}$ corresponds to the uncertainty from the \ion{O}{i} wavelength calibration, with a value of 0.2 km s$^{-1}$, $\sigma_{xy}$ represents the median uncertainty in the measured wavelength, computed across all spatial pixels (i.e., both x and y image coordinates) and across the whole FoV of the three rasters, i.e., encompassing all the data acquired during the three IRIS scans. Finally, $\sigma_{f}$ represents the offset between the FUVS and FUVL detectors. 
\end{appendix}
\end{document}